\documentstyle[prb,epsfig,aps]{revtex}
\input epsf

\newcommand{\centps}[2]{
        \begin{center}
                \epsfig{file=#1,height=#2mm}
        \end{center}
}

\newcommand{\twofig}[3]{
        \begin{center}
                \epsfig{file=#1,height=#3truecm}
                \epsfig{file=#2,height=#3truecm}
        \end{center}
}

\newcommand{\threefig}[4]{
        \begin{center}
                \epsfig{file=#1,height=#4truecm}
                \epsfig{file=#2,height=#4truecm}
                \epsfig{file=#3,height=#4truecm}

        \end{center}
}

\newcommand{\fourfig}[5]{
        \begin{center}
            \mbox{
                \setlength{\epsfxsize}{#5truecm}
                \setlength{\epsfysize}{#5truecm}
                \epsfbox{#1}
                \setlength{\epsfxsize}{#5truecm}
                \setlength{\epsfysize}{#5truecm}
                \epsfbox{#2}}
           \mbox{
                \setlength{\epsfxsize}{#5truecm}
                \setlength{\epsfysize}{#5truecm}
                \epsfbox{#3}
                \setlength{\epsfxsize}{#5truecm}
                \setlength{\epsfysize}{#5truecm}
                \epsfbox{#4}}
        \end{center}
}

\begin{document}
\draft

\twocolumn[\hsize\textwidth\columnwidth\hsize\csname @twocolumnfalse\endcsname

\title{Andreev Level Spectroscopy and Josephson Current Switching in a
3-Terminal Josephson Junction}

\author{H. Tolga Ilhan, H. Volkan Demir, and Philip F. Bagwell \\
Purdue University, School of Electrical Engineering \\ 
West Lafayette, Indiana 47907}

\date{\today}
\maketitle

\begin{abstract}

We calculate the electrical currents through a superconductor -
insulator - superconductor junction which is also weakly coupled to a
normal metal side probe.  The voltage $V$ applied to the normal metal
terminal controls the occupation of Andreev energy levels $E_n$, and
therefore controls the Josephson current flowing through these levels.
Whenever the probe voltage crosses an Andreev level, the Josephson
current changes abruptly by an amount equal to the current flowing
through the Andreev level. The differential conductance along the
normal metal terminal permits spectroscopy of the Andreev levels.  In
a short junction $(L \ll \xi_0)$, the critical current switches
abruptly from the Ambegaokar-Baratoff value to zero when the probe
voltage is approximately equal to the superconducting energy gap
($|eV| \simeq \Delta$).  The magnitude of the Josephson current
switching in a long junction $(L \gg \xi_0)$ , and the range of probe
voltages over which the Josephson current differs from its equilibrium
value, are much smaller than for three-terminal ballistic
superconductor - normal metal - superconductor junctions. 

\end{abstract}


submitted to Physical Review B-1

]  \narrowtext
\flushbottom
\section{Introduction}
\indent

The Andreev energy levels~\cite{andreev} in superconductor - normal
metal - superconductor (SNS) or superconductor - insulator -
superconductor (SIS) junctions, through which a large fraction of the
the Josephson current
flows~\cite{kulik,ishii,bardeen,kulikp,bvh,furusaki,vwees,bagwell},
are only weakly held in thermodynamic equilibrium with the two
superconducting contacts of a Josephson junction.  When the
quasi-particle energy is inside the superconducting gap,
quasi-particles cannot transmit into the superconductor, and also
cannot be injected into the Andreev levels from the
superconductor~\cite{vwees}. The coupled electron-like and hole-like
quasi-particles which form the Andreev levels orbit in continuous
periodic motion inside the normal (or insulating) region of the SNS or
SIS junction.  Quasi-particles in the Andreev levels are then
essentially thermodynamically isolated from the superconductors, yet
carry a large fraction of the supercurrent.  Only inelastic scattering
inside the superconductor forces the occupation factor for the Andreev
energy levels towards the equilibrium Fermi occupation factor of the
superconducting contacts.

An additional normal metal contact coupled to the Josephson junction
can directly inject quasi-particles into the Andreev energy levels
through elastic scattering processes, and therefore can directly
control the occupation of the bound levels.  The additional normal
metal probe coupled to the Josephson junction, shown in
Fig.~\ref{fig:geom}, models either a scanning tunneling microscope tip
or the gate electrode of a three-terminal Josephson
junction~\cite{morpurgo,schaepers,chang,wendinan,volkov,wilhelm,ilhan4T}.
Since the rate at which the superconducting contacts inject
quasi-particles into the Andreev levels through inelastic processes is
extremely small, even a normal metal probe only weakly in contact with
the Josephson junction will inject quasi-particles into the Andreev
levels much faster than the superconducting electrodes.  As long as
the normal metal probe is only weakly in contact with the Josephson
junction through a tunnel barrier, the main effect of the probe is
therefore to fix the occupation factor of the Andreev levels, leaving
the wavefunctions of the Andreev levels essentially unchanged from the
isolated Josephson junction.~\cite{vwees,chang,wendinan} The normal
metal probe therefore forces the effective Fermi level of the bound
Andreev states towards the Fermi level of the probe, rather than the
Fermi level of the superconductors.

\begin{figure}
\centps{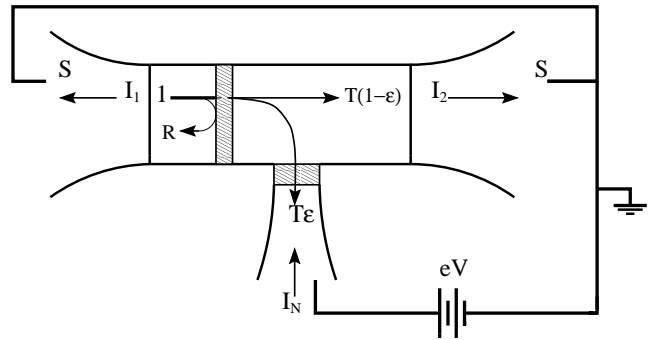}{45}
\caption{Josephson junction coupled to a normal metal side probe.  The
probe is biased at a voltage $V$ with respect to the two
superconductors. This normal metal probe controls the occupation
factor of the bound Andreev levels, since
quasi-particles having energies inside the superconducting gap
cannot transmit into either superconductor.}
\label{fig:geom}
\end{figure} 

Controlling the Andreev bound state occupation through a normal
terminal leads both to an abrupt switching of the Josephson current
$I(V)$, and a peak in the differential conductance $dI_N(V)/dV$ along
the normal terminal, whenever the probe voltage $V$ is equal to the
energy of an Andreev level.~\cite{chang} As a bound level is populated
or depopulated by the probe voltage, the Josephson current changes by
an amount equal to the current carried by the bound level.  As the
Andreev level is being filled by the probe, the differential
conductance $dI_N/dV$ along the normal metal lead also has a peak.
The density of Andreev levels in the SNS junction can therefore be
detected by measuring the differential conductance $dI_N/dV$ along the
normal metal lead. Together with a phase - biasing network of large
area Josephson junctions, this differential conductance spectroscopy
can be used to directly map~\cite{ouboter} the energy - phase relation
$E_n(\phi)$ and current - phase relation $I_n(\phi)$ of the
Andreev energy levels in different types of Josephson junctions.

One can obtain a more quantitative understanding of the Josephson
current switching and Andreev level spectroscopy in any Josephson
junction in the limit where the normal metal probe is weakly coupled
to the junction.  The total Josephson current $I(\phi, V)$ is a sum of
the currents flowing through discrete energies inside the
superconducting energy gap $I_{d}$ and through the continuum of energy
levels outside the gap $I_{c}$ as $I(\phi, V) = I_{d}(\phi, V) +
I_{c}(\phi)$.  Here $\phi= \phi_2 - \phi_1$ is the superconducting
phase difference.  The contribution $I_{c}$ to the Josephson current
by scattering states outside the superconducting energy gap is
essentially unchanged by the probe voltage $V$, since the
superconducting contacts can easily inject quasi-particles into
Andreev resonances in the energy continuum.  Scattering states outside
the energy gap therefore remain in equilibrium with the
superconducting contacts.  The portion of the Josephson current
$I_{d}$ flowing through the Andreev levels, however, is~\cite{chang}
\begin{equation}
\matrix{
I_{d}(\phi, V) = \sum_n \{ I_n^{-}(\phi) f(E_n^{-}(\phi)-eV) \;\; + 
\vspace{0.05in} \cr
\hspace{1.0in}I_n^{+}(\phi) f(E_n^{+}(\phi)-eV) \}}
\; .
\label{Idiscretebody}
\end{equation}
In Eq.~(\ref{Idiscretebody}) the $I^{\pm}_{n}(\phi) =
(2e/\hbar)(dE^{\pm}_{n}(\phi)/ d\phi)$ are the currents carried by
`forward' and `reverse' Andreev levels $E^{\pm}_{n}$ before adding the
side probe.  The probe voltage $eV$ appears inside the Fermi factors
$f$ in Eq.~(\ref{Idiscretebody}) as an effective electrochemical
potential for the Andreev levels.  Therefore,
Eq.~(\ref{Idiscretebody}) implies that the contribution of each
Andreev level to the total Josephson current can be switched on or off
by varying the probe voltage $V$.

The tunneling current through the normal metal probe measures the
local density of quasi-particle states in the Josephson junction.  It
is well known from tunneling spectroscopy of normal metals that the
tunneling current is proportional to the local density of states at
the surface~\cite{tunnelspec}, and this also holds true for
superconducting tunnel junctions. The tunneling spectroscopy of
Andreev levels in a Josephson junction, using tunneling current from
the normal probe then, corresponds to
\begin{eqnarray}
\frac{dI_N}{dV} = \frac{4e^2}{h} \sum_{n,\alpha}
\left( \frac{\Gamma_n^2}{\Gamma_n^2 + (eV-E_n^\alpha)^2} \right)
\; ,
\label{didv}
\end{eqnarray}
when $|eV| < \Delta$. Here $\Gamma_n$ is the width of Andreev level
$n$, which is proportional to the coupling constant $\epsilon$.
Although Ref.~\cite{chang} derived
Eqs.~(\ref{Idiscretebody})-(\ref{didv}) only for a ballistic SNS
junction, they should describe any type of Josephson junction.  The
numerical simulations we present in the following sections follow from
the scattering theory in Appendix A, and can be understood using
Eqs.~(\ref{Idiscretebody})-(\ref{didv}).

In this paper we consider the Josephson current switching $I(V)$ and
differential conductance spectroscopy of the Andreev levels
$dI_N(V)/dV$ as we vary the voltage $V$ along the normal terminal in a
three-terminal SIS junction.  The details of the Josephson current
switching and spectroscopy of the Andreev states in a three-terminal
SIS junction differ considerably from the ballistic SNS
junction~\cite{chang}.  In a short SIS junction, having $(L \ll
\xi_0)$, the presence of an insulator forces the Andreev energy levels
to the edge of the superconducting energy gap.  The Josephson current
in a short SIS junction therefore switches to zero when the voltage on
the normal terminal is approximately equal to the energy gap, i.e. $eV
\simeq \Delta$, as we discuss in section 3. Section 4 shows that
the terminal I-V characteristics of SIS Josephson junctions longer
than the healing length ($L \gg \xi_0$) are more complex than those of
short SIS junctions. The size of the nonequilibrium Josephson current,
the regularity of its switching behavior, and the voltage range over
which the terminal currents are constant, are sensitive both to the
barrier transmission $T$ and the position of the tunnel barrier in the
junction. We also discuss the special limiting case where the long SIS
junction has inversion symmetry~\cite{wendin}.

\section{Short Josephson Junction}
\indent

In a short SNS junction ($L \ll \xi_{o}$), the Josephson current
flowing into either superconductor switches on or off as we vary the
bias voltage on the side probe. It is well known both that short
Josephson junctions contain only two Andreev levels, and that all the
Josephson current flows through these levels ($I_c = 0$). Depopulating
(or populating) both levels therefore forces the Josephson current to
zero. Consider the SIS junction having transmission probability $T =
2.5\%$ in Fig.~\ref{fig:shortphase}. The two Andreev levels
$E_n(\phi)$ for the SIS junction are shown in
Fig.~\ref{fig:shortphase}(a). The horizontal lines in
Fig.~\ref{fig:shortphase}(a) correspond to bias voltages near the
energy gap, namely $eV = \pm 0.995 \Delta$.  For positive bias
voltages, injection of a quasi-particle from the normal metal terminal
fills any Andreev level having energy less than $E_n \leq eV$.  When
only the lowest Andreev level satisfies $E_n \leq eV$, only the lowest
energy level contains a quasi-particle. Consequently, when $eV \leq
\sqrt{R} \Delta$,\cite{bagwell} the Josephson current through the SIS
junction is nearly the same as without the probe, i.e. we recover the
standard Ambegaokar-Baratoff
result~\cite{bagwell,ambegaokar,haberkorn,zaitsev,arnold,glazman,furusakiBC,been1},
as shown in Fig.~\ref{fig:shortphase}(b).

\begin{figure}
\twofig{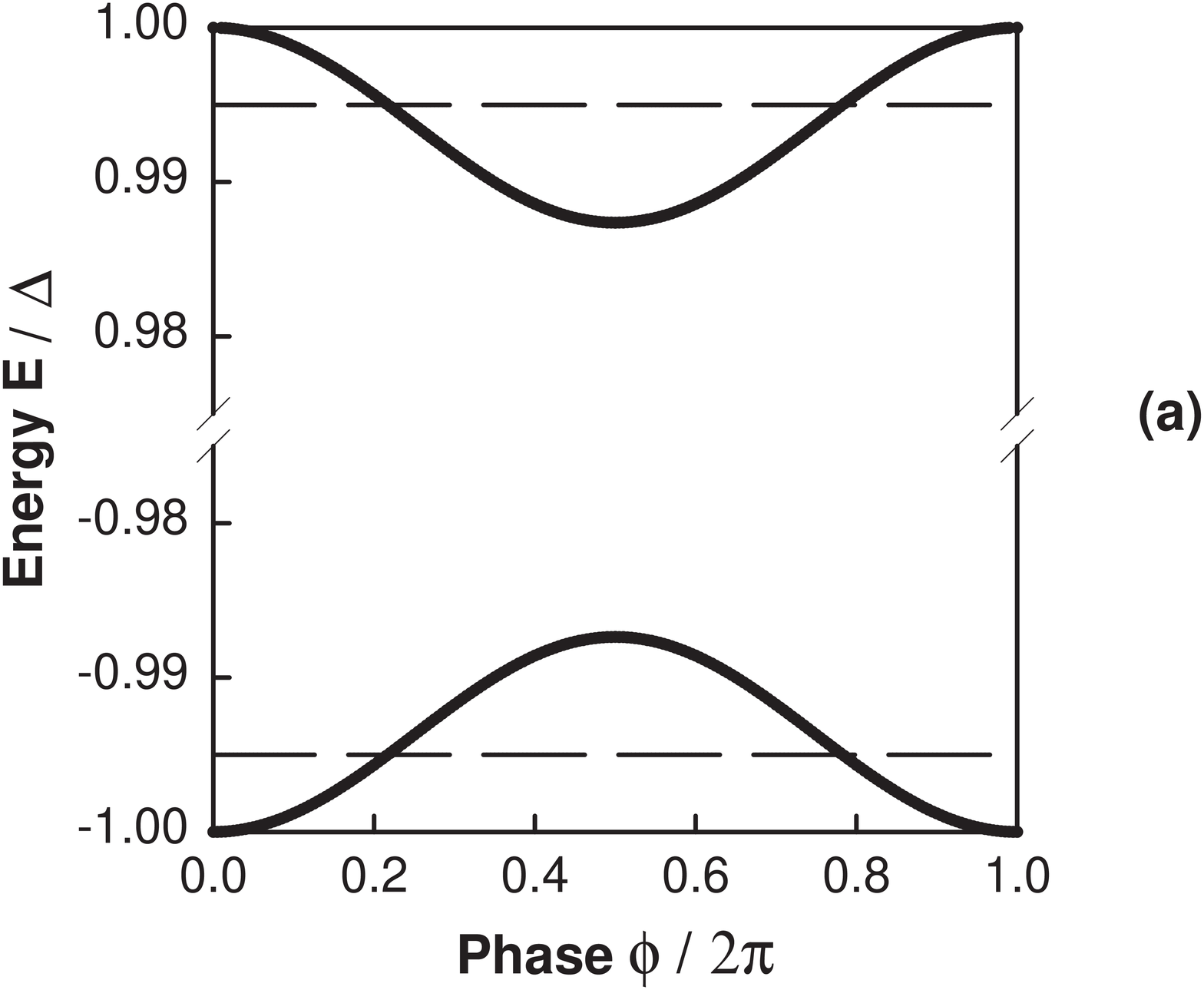}{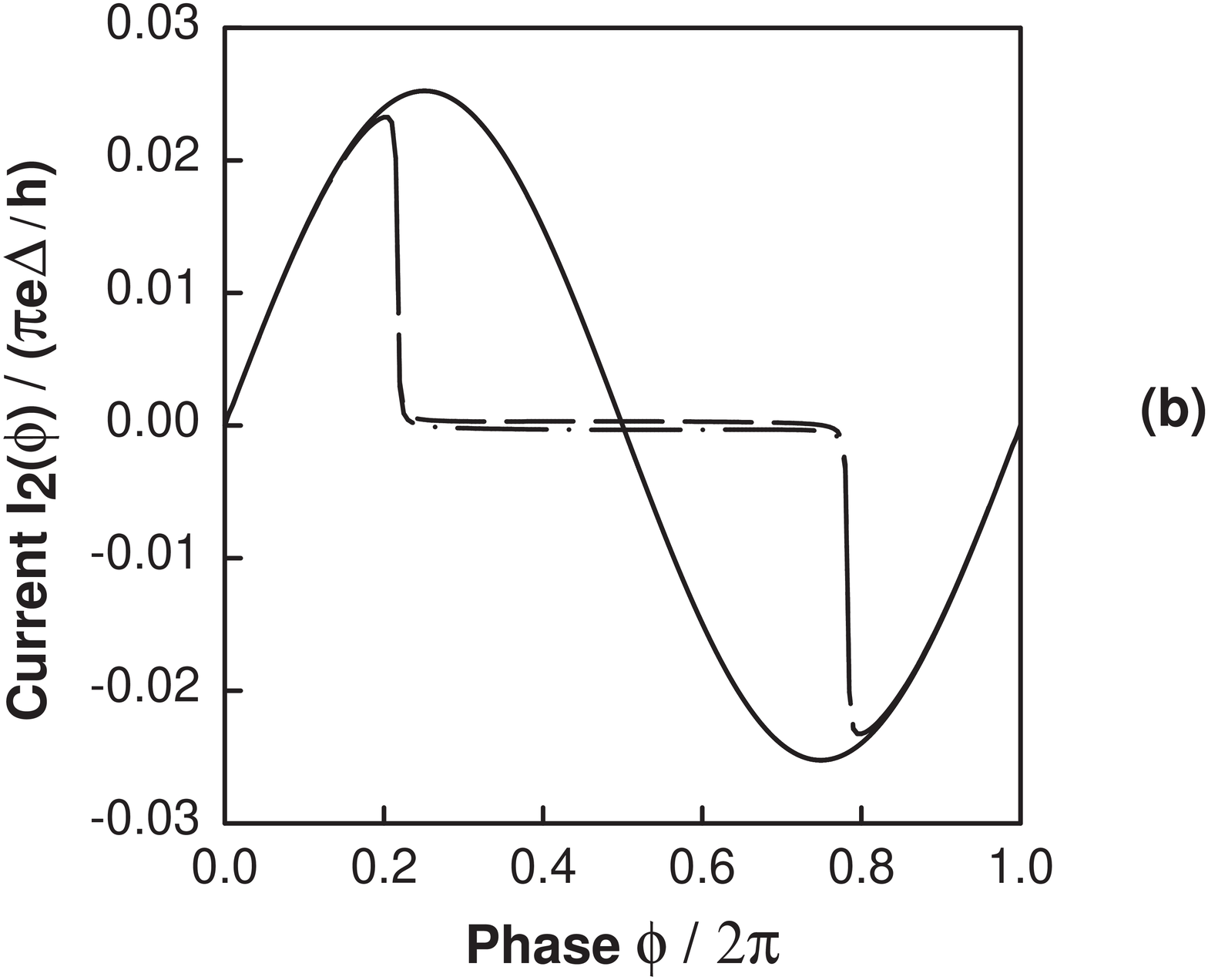}{5.5}
\caption{(a) Andreev levels $E^{\pm}(\phi)$ (solid)
and bias voltage eV = $\pm0.995\Delta$ (dashed) in a short SIS
junction. (b) The probe voltage crossing an Andreev level forces the
Josephson current to zero, except for a small leakage current.}
\label{fig:shortphase}
\end{figure} 

By varying the gate voltage, we can switch the Josephson current in a
short SNS junction on or off.  At a fixed phase difference, the two
Andreev levels carry equal amounts of current but in opposite
directions.  Therefore, when both Andreev levels are filled (or empty)
the Josephson current is nearly zero as shown in
Fig.~\ref{fig:shortphase}(b).  We also conclude that the Josephson
current-phase relation is nearly the same whether the side probe has a
negative or positive bias voltage.  The small difference between the
two Josephson currents for $eV = \pm 0.995 \Delta$ shown in
Fig.~\ref{fig:shortphase}(b) is due to the small leakage current from
the gate. The leakage current is small because the coupling strength
$\epsilon = 0.1\%$ in Fig.~\ref{fig:shortphase}. The small leakage
current implies that, as we vary the gate voltage between $\sqrt{R}
\Delta \leq eV \leq
\Delta$, the Josephson current switches from the Ambegaokar-Baratoff
value to approximately zero.

Reducing the insulator transmission $T$ forces the switching voltage
and differential conductance peak towards the energy gap, i.e. $eV =
\Delta$, as shown in Fig.~\ref{fig:shortswitch}. 
In Fig.~\ref{fig:shortswitch} the transmission probability decreases
from $T = 100\%, 36\%, 12\%$, the probe coupling is weak ($\epsilon =
0.1\%$ in (a)-(b) and $\epsilon = 5\%$ in (c)), and we fix $\phi = 0.3
(2\pi)$ (b)-(c). With decreasing transmission $T$, an energy gap opens
and forces the Andreev levels to the superconducting gap edge in
Fig.~\ref{fig:shortswitch}(a).  Lowering the transmission coefficient
in SIS junctions, therefore forces the switching voltage and
differential conductance peak towards the superconducting gap at $eV =
\Delta$, as shown in Fig.~\ref{fig:shortswitch}(b)-(c), respectively.

\begin{figure}[tbh]
\threefig{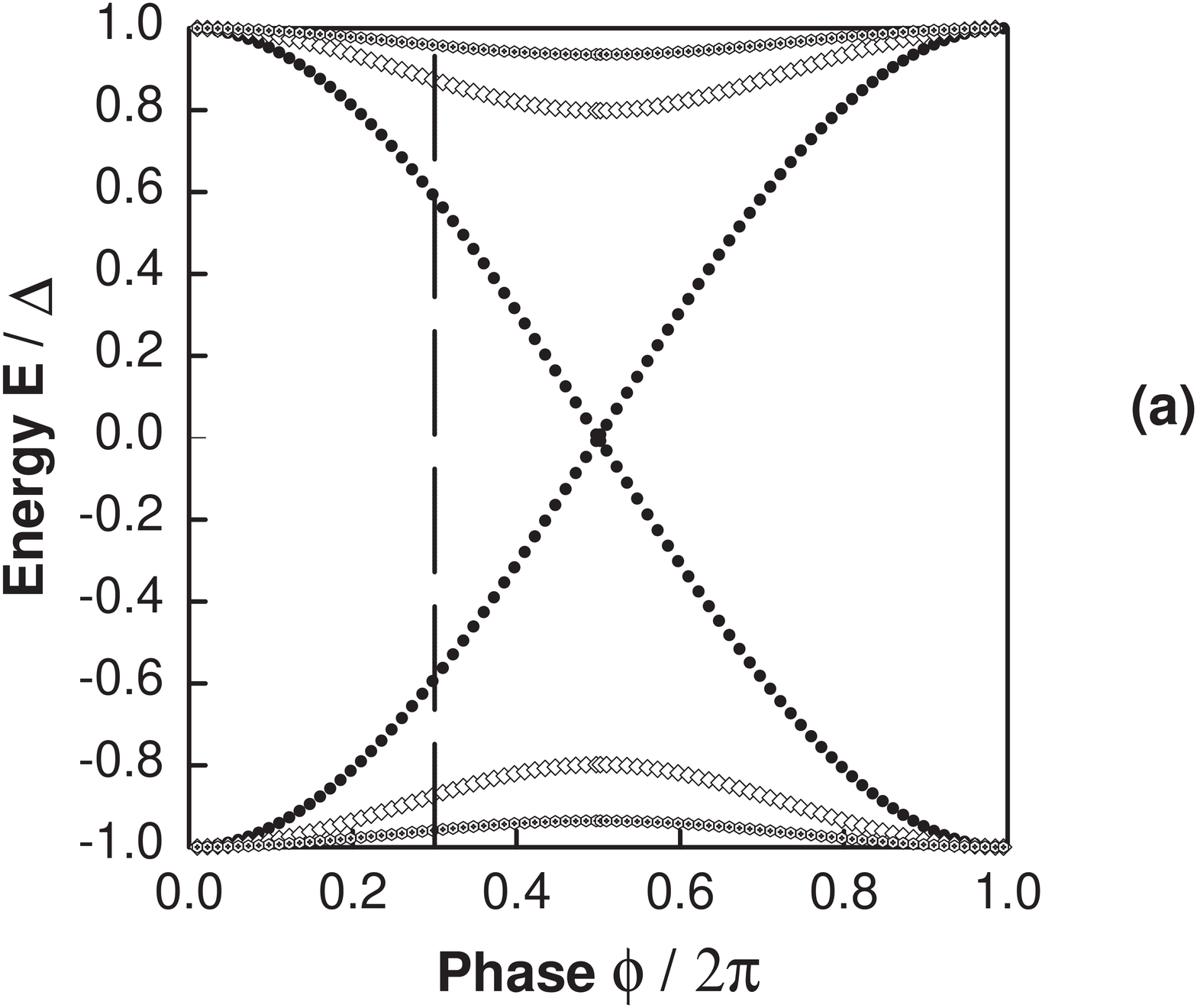}{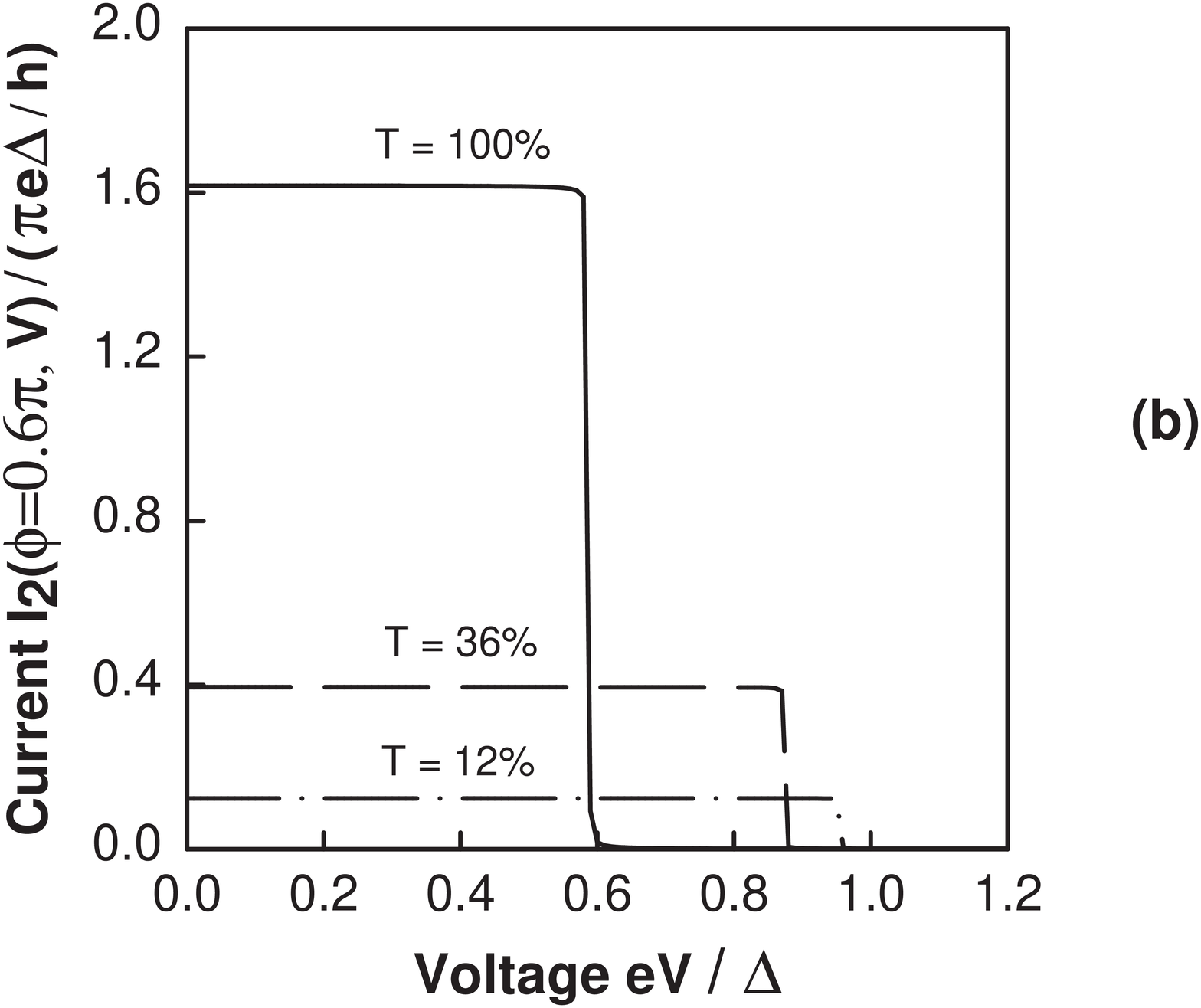}{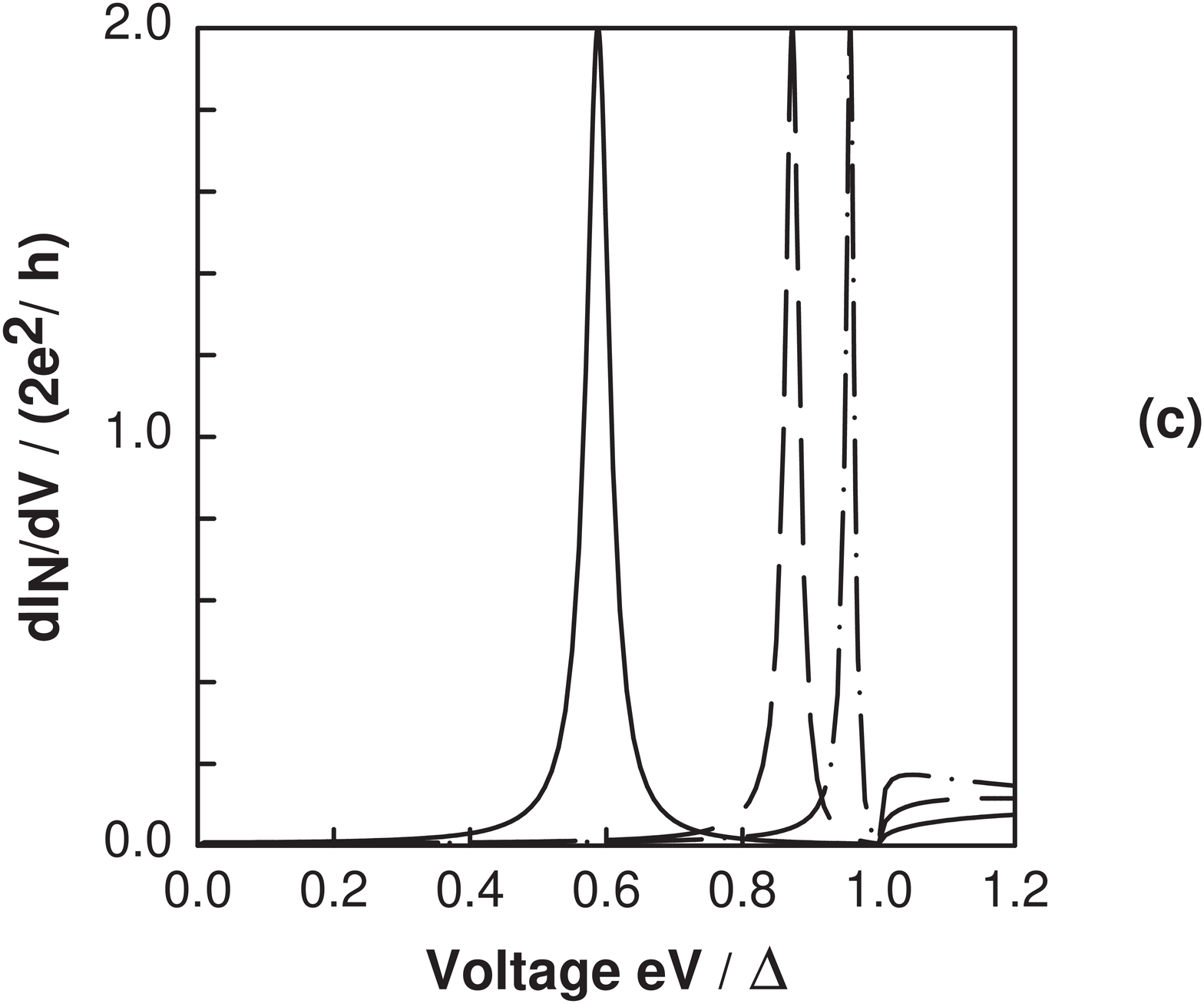}{5.5}
\caption{(a) Andreev levels $E^{\pm}(\phi)$, (b) Josephson
current $I(\phi,V)$, and (c) differential conductance $dI_{N}/dV
(\phi,V)$ along the normal metal probe.  Decreasing transmission
probability ($T = 100\%, 36\%, 12\%$) forces both the switching
voltage in (b) and peak in the differential conductance in (c) towards
the gap edge.}

\label{fig:shortswitch}
\end{figure}

\twocolumn[\hsize\textwidth\columnwidth\hsize\csname @twocolumnfalse\endcsname
\begin{figure}
\fourfig{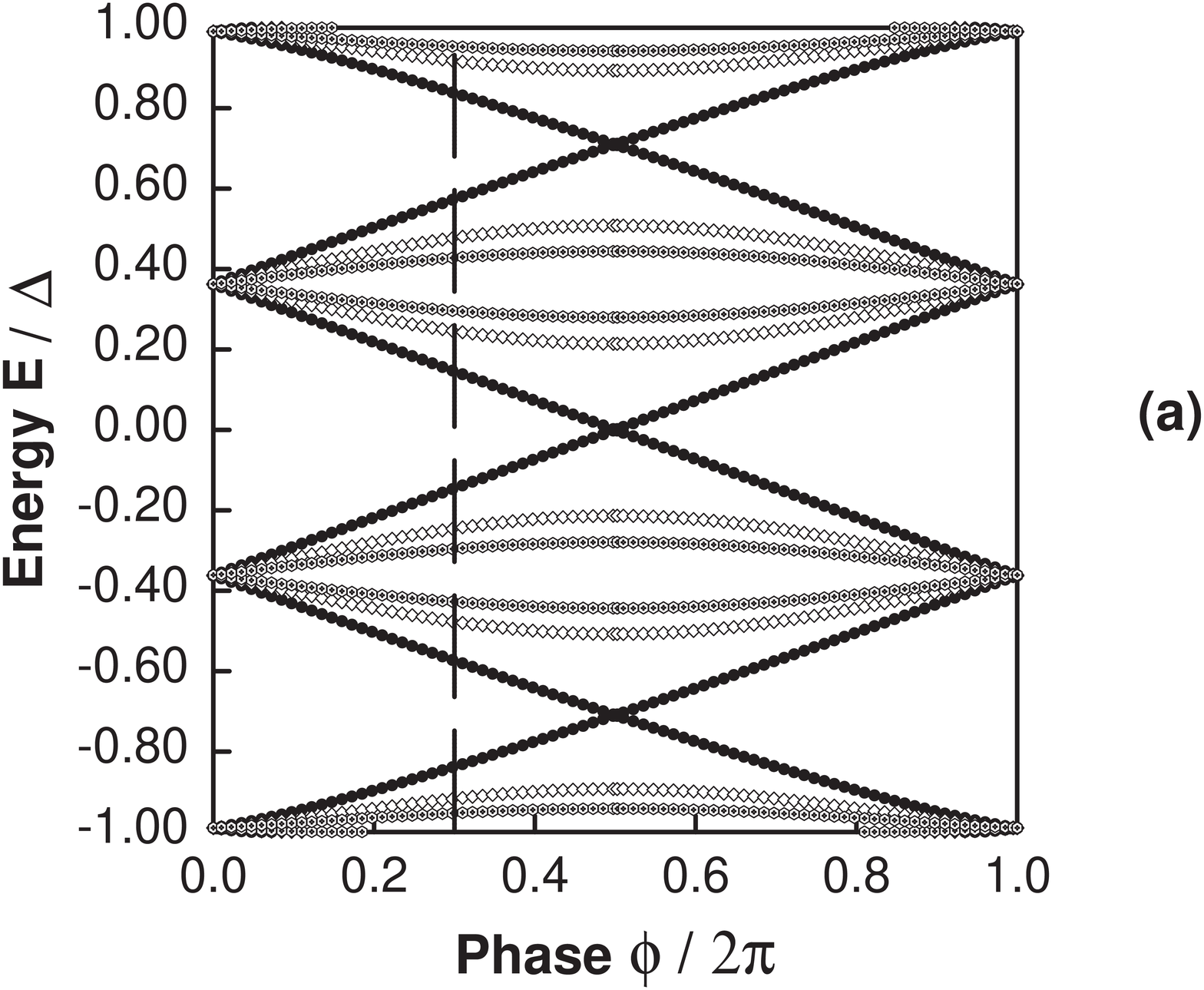}{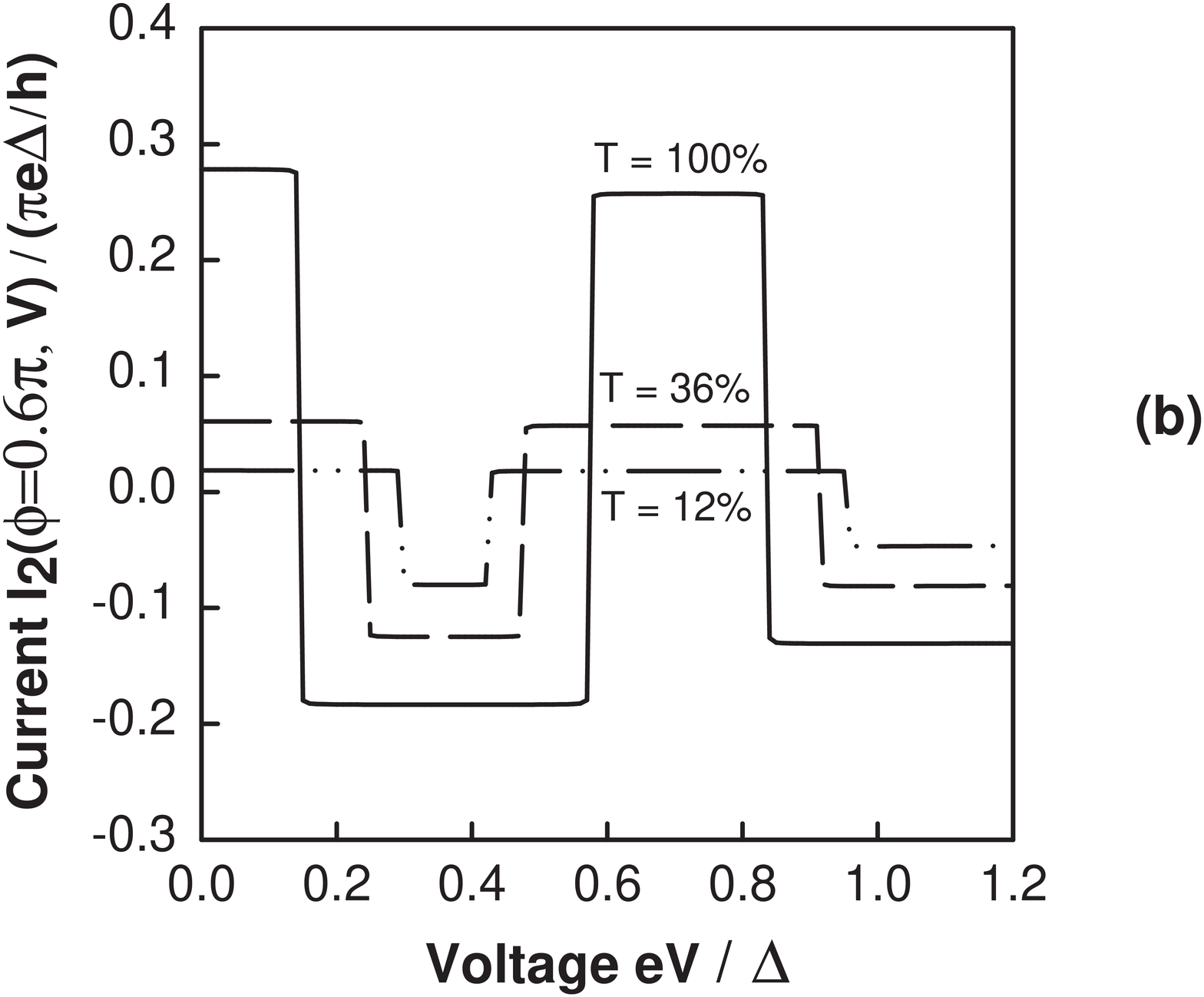}{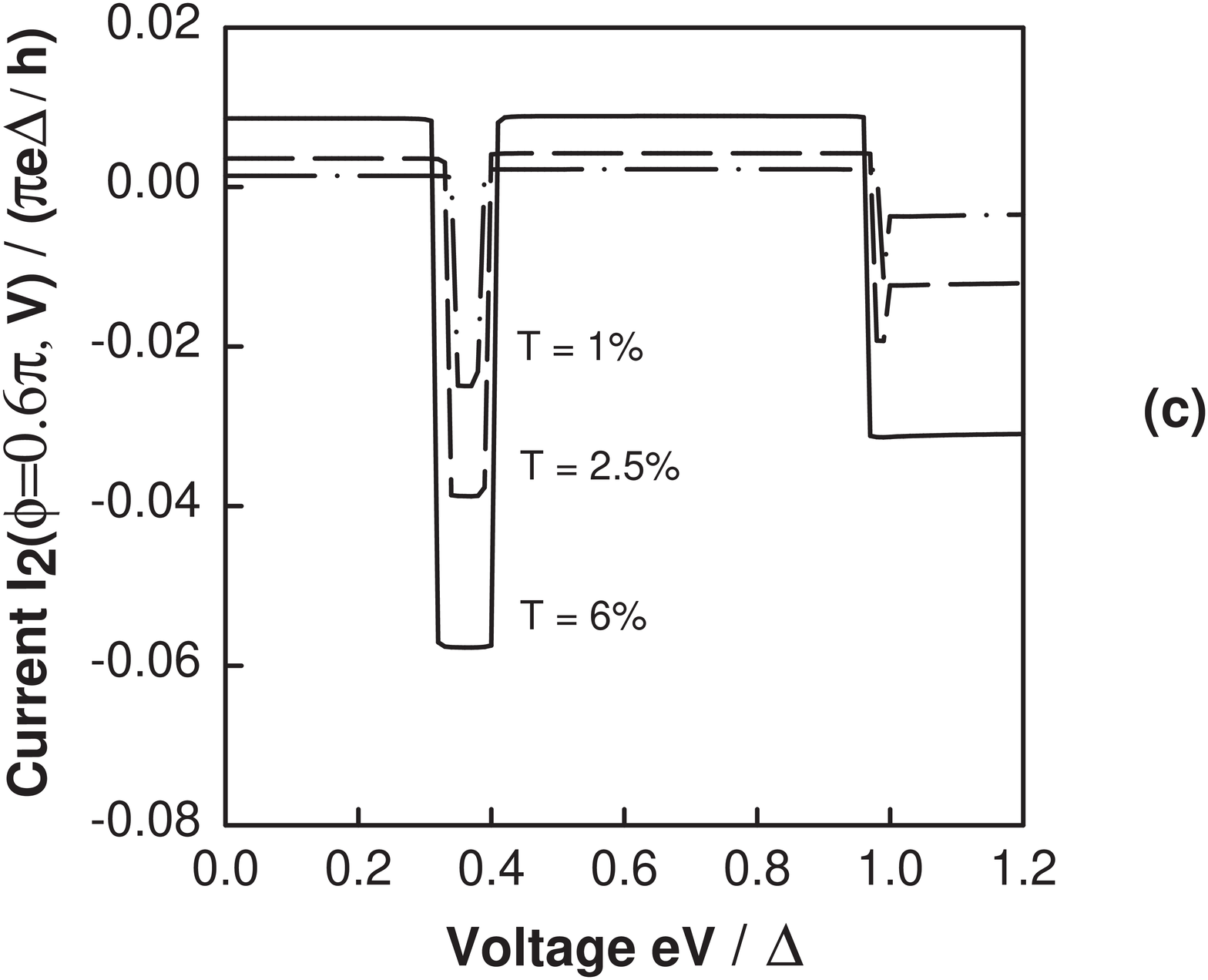}{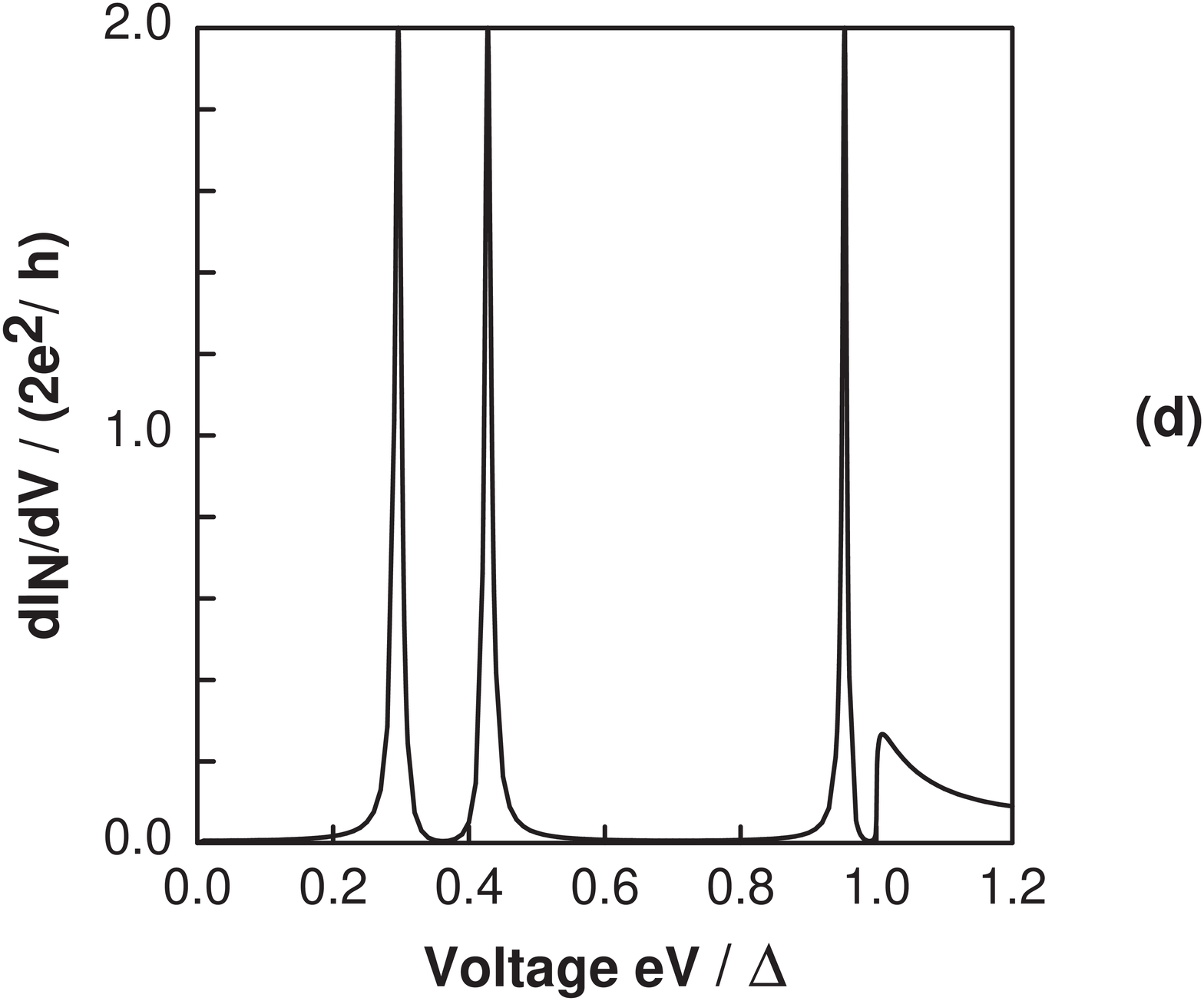}{6.5}
\caption{A long $(L > \xi_{0})$ SIS junction which has inversion symmetry.
(a) Andreev levels, (b)-(c) Josephson currents, and (d) differential
conductance along the normal probe. With decreasing barrier
transmission, the nonequilibrium Josephson current exceeds the
equilibrium current ($V=0$) over a narrow voltage range.  Ballistic
junctions ($T=100\%$) have the largest currents both in and out of
equilibrium.}

\label{fig:longsym}
\end{figure} 

] \narrowtext

We can understand why the Josephson current in
Fig.~\ref{fig:shortswitch}(b) is constant before switching off when
the bias voltage crosses an Andreev level at $E_n = eV$ by considering
the Andreev level structure in Fig.~\ref{fig:shortswitch}(a).  For a
positive bias voltages and fixed phase difference $\phi$, the lowest
Andreev level ($E_n \leq 0$) carries the total current until the upper
level crosses $E_n = eV$.  The zero temperature Josephson current is
therefore unchanged for voltages smaller than $|eV| \leq |E_n|$.
Decreasing $T$ reduces the Josephson current in
Fig.~\ref{fig:shortswitch}(b), since the lowest Andreev level carries
a smaller current with smaller transmission $T$.  The differential
conductance $dI_N/dV$ along the probe also has a peak whenever the
bias voltage crosses a new Andreev level, as shown by comparing
Fig.~\ref{fig:shortswitch}(b)-(c).

\section{Long Josephson Junction}
\indent

Spectroscopy of the Andreev levels and the Josephson current switching
as a function of the gate voltage $V$ change significantly when the
junction length $L$ becomes comparable to the BCS healing length
$\xi_0$. The number of levels is proportional to $L/\xi_0$, so more
Andreev levels $E_n(\phi)$ become bound in the pair potential well.
In addition, the interference pattern between quasi-particle waves
multiply reflected between the NS interface and the tunnel barrier
depends on the junction length $L/\xi_0$ and the symmetry of the SIS
junction~\cite{wendin}. The current $I_n(\phi)$ flowing through the
Andreev levels also depends on the junction length and symmetry of the
scattering potential. To illustrate the variation of the Josephson
current and probe current with these parameters, we study in this
section a `symmetric' SIS junction (where the impurity is in the
middle of the normal metal region) and an `asymmetric' junction
lacking this inversion symmetry.

\subsection{Symmetric Junction}
\indent
It has long been known that the net equilibrium Josephson current in
long SNS and SIS junctions is a small difference between much larger
positive and negative currents flowing in
equilibrium~\cite{ishii,bardeen,svidzinski,furusaki,vwees,bagwell}.
For a Josephson junction in equilibrium, adding these large
counterflowing currents produces a net current proportional to the
barrier transmission $T$. Ref.~\cite{wendin} quantified the magnitude
of the larger electrical currents (which cancel in equilibrium).  In
long, low transmission Josephson junction having inversion symmetry
Ref.~\cite{wendin} showed that these larger currents are proportional
to $\sqrt{T}$.  Ref.~\cite{wendin} also suggested that one could probe
these `giant' currents by inducing a non-equilibrium population of the
Andreev levels.

Fig.~\ref{fig:longsym} shows the (a) Andreev levels, (b)-(c) Josephson
currents, and (d) differential conductance along the normal probe in a
long $(L = 6.6 \xi_{0})$, symmetric SIS junction. The probe coupling
in Fig.~\ref{fig:longsym} is also weak, so that $\epsilon = 0.1\%$ in
(a)-(c) and $\epsilon = 5\%$ in (d). The Andreev levels in
Fig.~\ref{fig:longsym}(a) do not split at $\phi = 0, 2\pi$, as the
transmission decreases from $T = 100\%$ (filled) to $36\%$ (empty) or
$12\%$ (dotted), as one might expect.~\cite{bagwell} The Andreev
levels in the symmetrical junction do not split at $\phi = 0, 2 \pi$
because of a geometrical symmetry in the junction.  Upon normal
reflection from the tunnel barrier, a quasi-particle cannot tell
whether it is on the left or right side of the barrier when $\phi = 0,
2
\pi$. Therefore, the energy levels at $\phi = 0, 2 \pi$ are unaffected
by the presence of a tunnel barrier. When the phase difference is not
$\phi = 0, 2 \pi$ this geometrical symmetry is broken, so the
degenerate energy levels do split at any phase value other than $\phi
= 0, 2 \pi$.  This failure of the energy levels to split at $\phi = 0,
2 \pi$ in s-wave SIS junctions is exactly the same geometrical
symmetry leading to the `midgap' energy levels in Josephson junctions
formed from d-wave superconductors~\cite{hu,riedel-dnd}.

To obtain a formula for the Andreev levels in
Fig.~\ref{fig:longsym}(a), we set $a = L/2$ in Eqs.~(7)-(8) of
Ref.~\cite{bagwell}.  The `effective phase' $\alpha$ in the symmetric
SIS junction then simplifies to
\begin{eqnarray}
\sin(\alpha) \simeq \alpha \simeq 2\sqrt{T}|\sin(\frac{\phi}{2})|,
\label{phase_bagwell}
\end{eqnarray}
leading to the Andreev levels 
\begin{eqnarray}
E_n^\pm \simeq \frac{\Delta \xi_0}{L+2\xi_0}
(2\pi n - \pi \mp 2\sqrt{T}|\sin(\frac{\phi}{2})|) \; .
\label{En-bag}
\end{eqnarray}
The geometrical symmetry also produces an additional resonant
enhancement of the Josephson current in each Andreev level. By setting
$a=L/2$ in Eqs.~(7), (15), and (16) of Ref.~\cite{bagwell}, we obtain
\begin{eqnarray}
I_n^{\pm} = \pm \frac{e v_F}{L + 2\xi(E_n^\pm(\phi))} \sqrt{T} 
\cos(\frac{\phi}{2}) \; .
\label{In-bag}
\end{eqnarray}
for symmetrical junction. In short junctions the variation of
$E_n^\pm(\phi)$ in Eq.~(\ref{In-bag}) produces an Andreev level
current proportional to $T$~\cite{bagwell}, while in long junctions
$E_n^\pm(\phi) \simeq {\rm constant}$ when the barrier transmission $T
\ll 1$. Consequently, $I_n \propto \sqrt{T}$ in long, low
transmission, symmetric Josephson junctions, again for the same
reasons as the resonant enhancement of Josephson current in $d$-wave
superconducting junctions.~\cite{riedel-dnd}.
Eqs.~(\ref{En-bag})-(\ref{In-bag}) are the same as found by Wendin and
Shumeiko in Ref.~\cite{wendin}.

Fig.~\ref{fig:longsym}(b)-(c) shows the Josephson current switching
for the long, low transmission, symmetric Josephson junction. (We fix
$\phi = 0.3 (2\pi)$ in Fig.~\ref{fig:longsym}(b)-(d).) The changes in
the Josephson current have equal magnitudes until the gate voltage
approaches the energy gap, similar to the ballistic SNS
junction~\cite{chang}. Unfortunately, this switching occurs only over
a much narrower range of gate voltages than in the ballistic SNS
junction. Comparing Figs.~\ref{fig:longsym}(b)-(c) we see that the
range of gate voltages where the Josephson current changes from its
equilibrium value becomes much narrower as the transmission decreases.
The $\sqrt{T}$ versus $T$ effect is also clearly visible in
Fig.~\ref{fig:longsym}(c), though none of the `giant' Josephson
currents~\cite{wendin} in Fig.~\ref{fig:longsym}(c) are as large as
the ballistic SNS junction in Fig.~\ref{fig:longsym}(b). The ballistic
SNS junction has both the largest equilibrium Josephson current and
the largest switching amplitude of the Josephson current with
variation in the gate voltage.

The range of voltages over which the Josephson current differs from
its equilibrium value in a ballistic SNS junction is approximately
half of the energy gap, as shown in Fig.~\ref{fig:longsym}(c).  In
contrast, the Josephson current in a long SIS junction with inversion
symmetry differs from its equilibrium value only over a very narrow
range of voltages, as also shown in Fig.~\ref{fig:longsym}(c). We can
infer from Eq.~(\ref{En-bag}) that the range of voltage over which the
`giant' Josephson current of Ref.~\cite{wendin} occurs is proportional
to $\sqrt{T} \Delta (\xi_0/(L+2\xi_0))$, namely the bandwidth of the
Andreev level.  Fig.~\ref{fig:longsym}(d), for the same junction
having transmission probability $T = 12\%$, shows the differential
conductance $dI_N/dV$ along the probe has a peak whenever the probe
voltage crosses a new Andreev level.

\subsection{Asymmetric Junction}
\indent

\begin{figure}

\fourfig{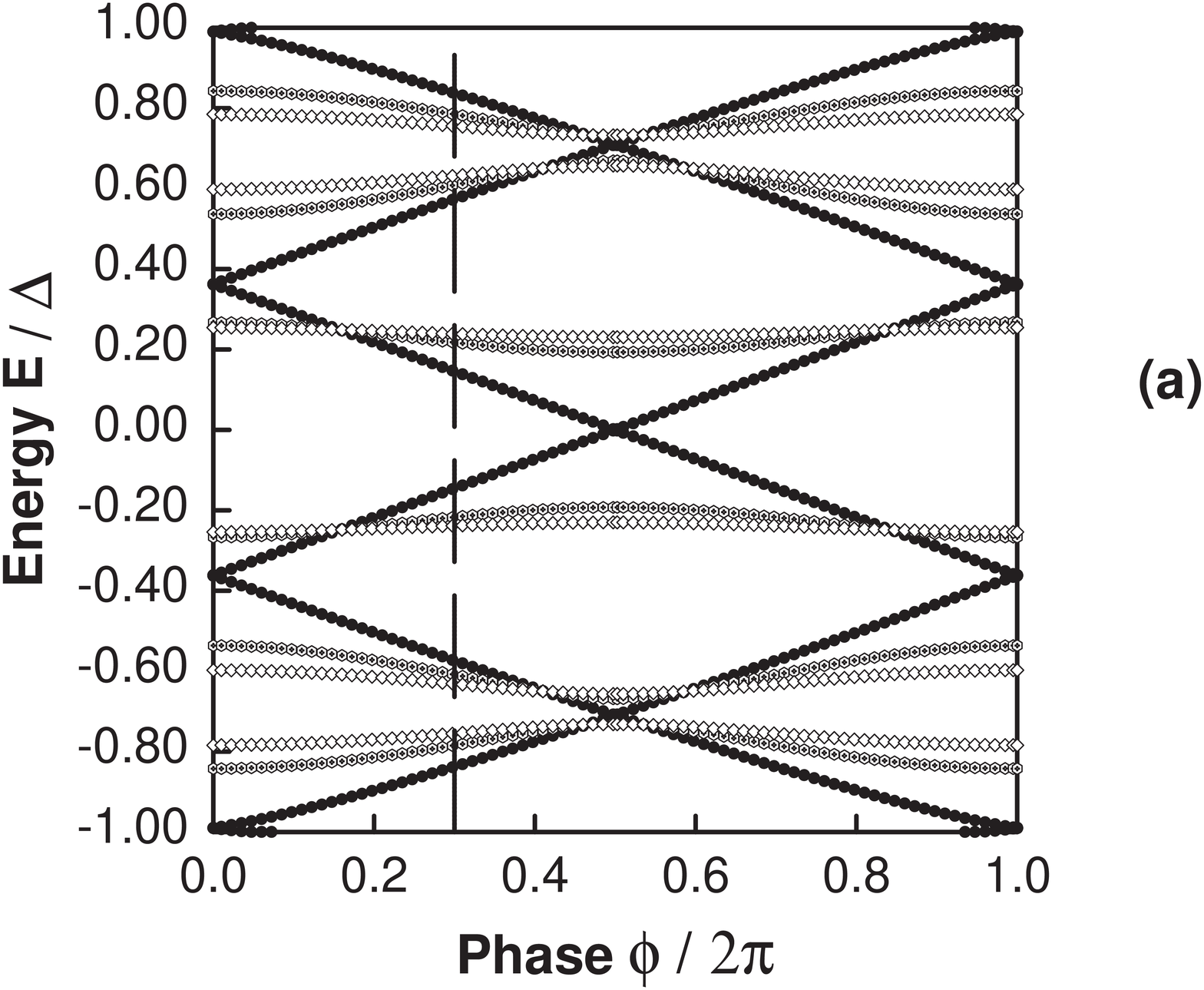}{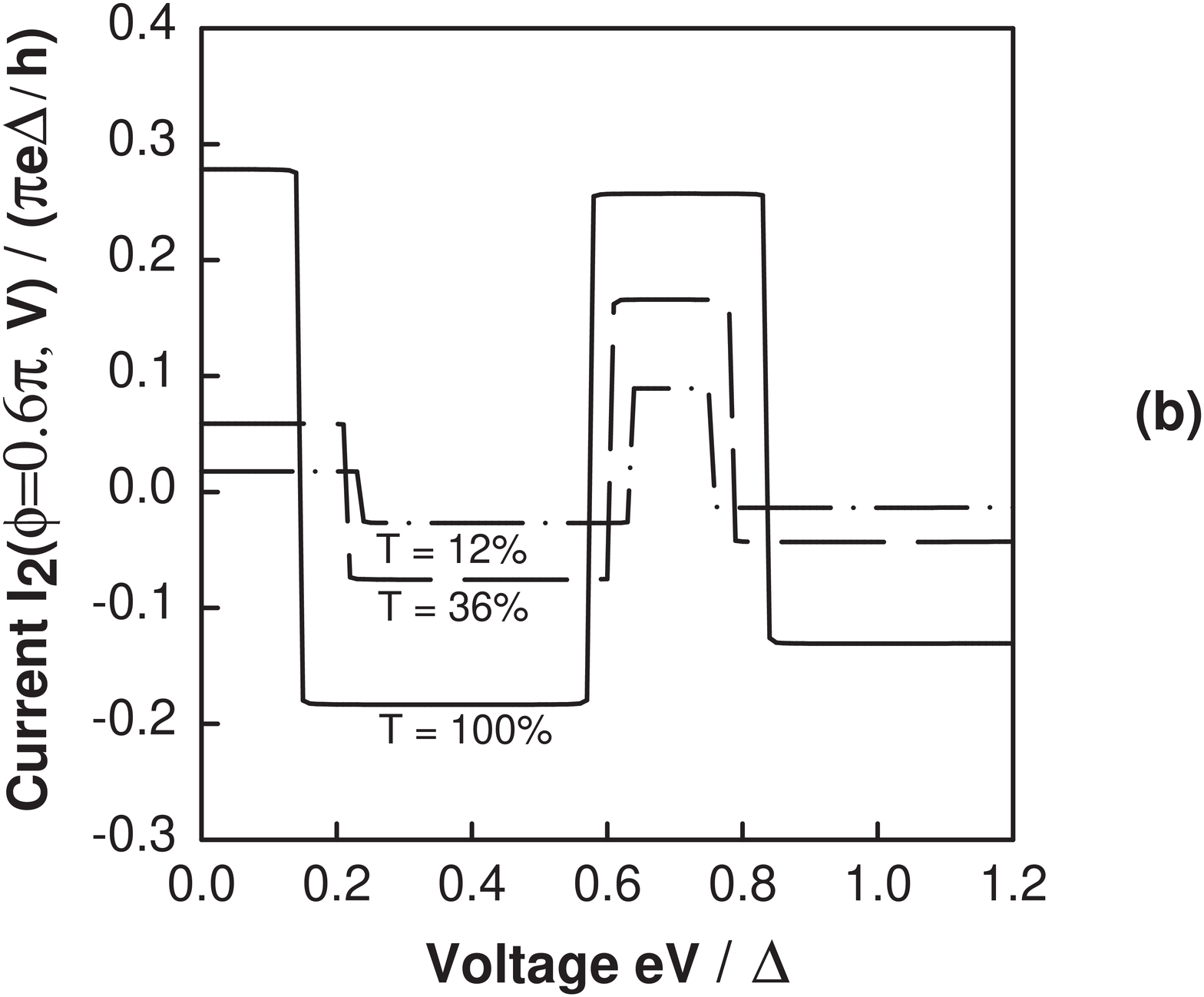}{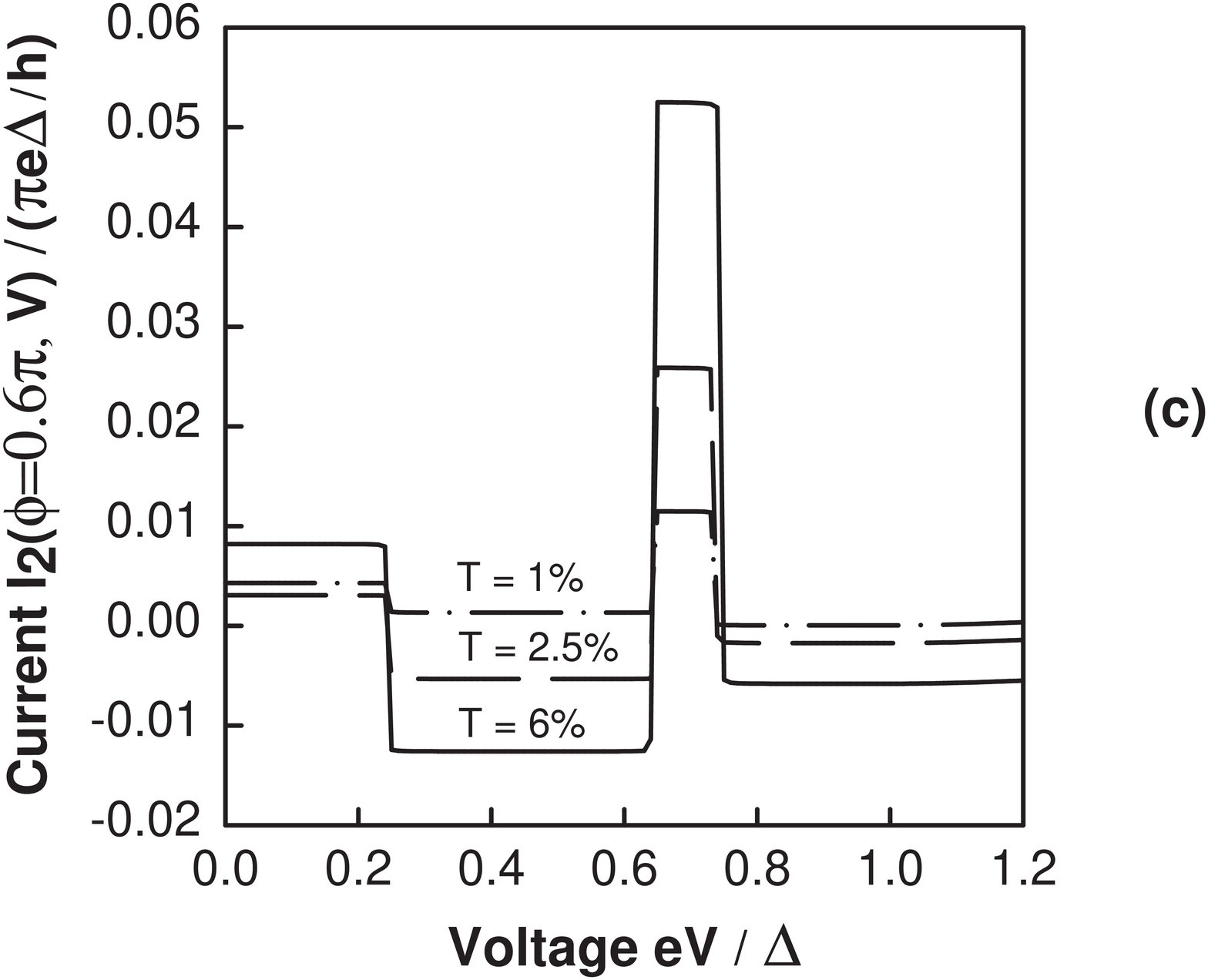}{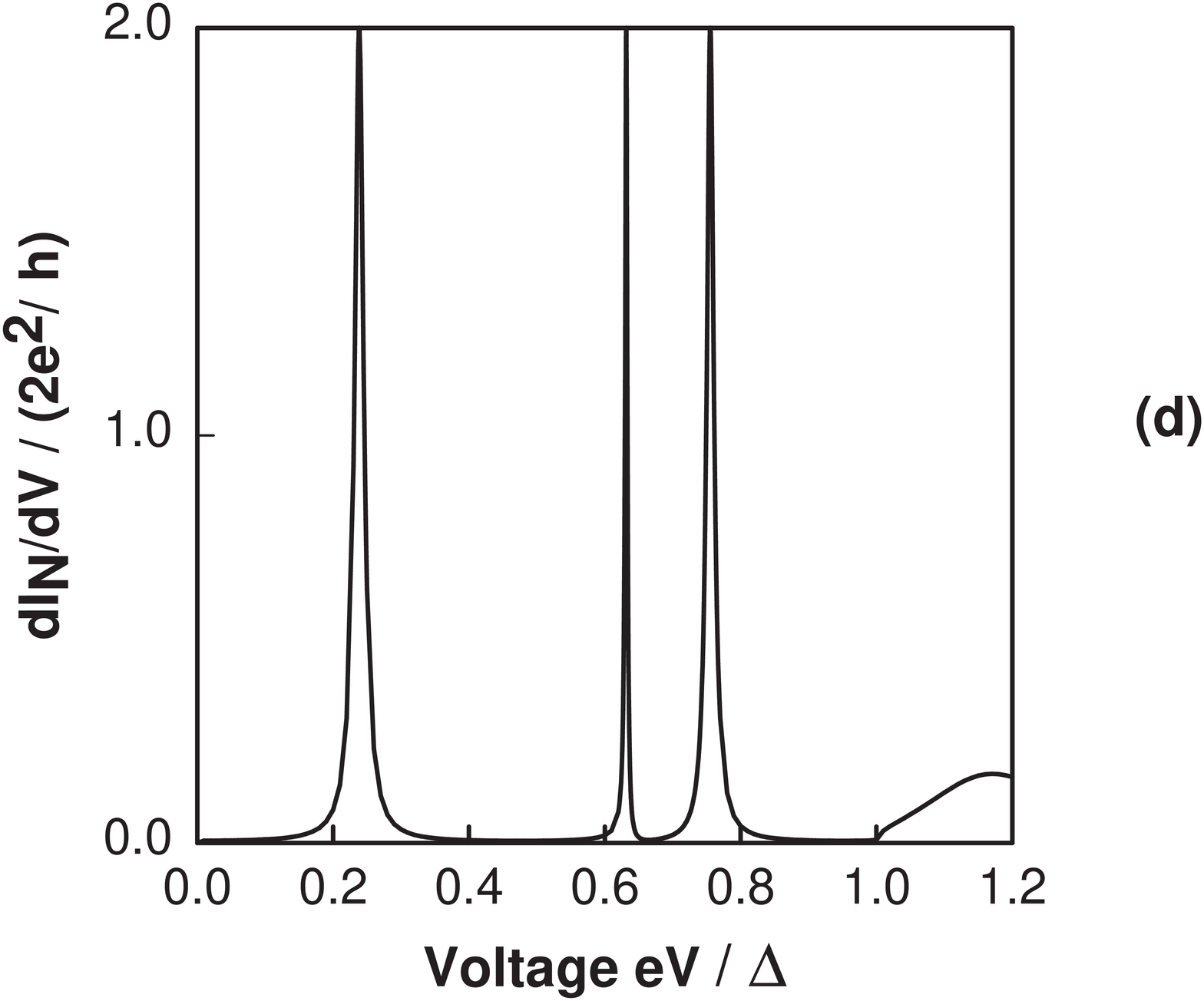}{6.5}
\caption{A long $(L > \xi_{0})$ SIS junction without inversion symmetry.
(a) Andreev levels, (b)-(c) Josephson currents, and (d) differential
conductance along the normal probe. A small Andreev energy gap leads
to large nonequilibrium Josephson currents, but over a narrow voltage
range.  Larger Andreev energy gaps produce a nonequilibrium current of
nearly the same magnitude as the equilibrium current, but over a wider
voltage range.  Ballistic junctions ($T=100\%$) again have the largest
currents, both in and out of equilibrium.}

\label{fig:longasym}
\end{figure}

Fig.~\ref{fig:longasym} shows the (a) Andreev levels, (b)-(c)
Josephson currents, and (d) gate current in an asymmetric junction
where $a = L/5$. The Andreev levels in Fig.~\ref{fig:longasym}(a)
correspond to a long SIS junction ($L = 6.6\xi_0$) where the
transmission probability $T = 100\%, 36\%, 12\% $ and the coupling
strength $\epsilon = 0.1\%$. The presence of an impurity removes all
of the degeneracy in the Andreev level spectrum in an asymmetric SIS
junction, though some energy levels split more than others. For
example, the energy gap  at $\phi = \pi$ and $|E|
\simeq 0.7 \Delta$ is much smaller than the other gaps. 

\twocolumn[\hsize\textwidth\columnwidth\hsize\csname @twocolumnfalse\endcsname

] \narrowtext

The small energy gaps present in the Andreev level spectrum, even in
asymmetric SIS junctions, produce Andreev level currents which
approach the `giant' $\sqrt{T}$ variation of the Josephson current of
Ref.~\cite{wendin}. Figs.~\ref{fig:longasym}(b)-(c) show the Josephson
current switching in a long SIS junction ($L = 6.6\xi_0$) having
coupling strength $\epsilon = 0.1\%$. We also fix the phase at $\phi =
0.3 (2\pi)$ in Figs.~\ref{fig:longasym}(b)-(d). For the small
transmission coefficients in Fig.~\ref{fig:longasym}(c), we see that
the equilibrium Josephson current decreases much more rapidly than the
current carried in the Andreev levels near $|E|
\simeq 0.7 \Delta$ (due to the small energy gap near  $|E|
\simeq 0.7 \Delta$). Comparing the asymmetric junction of 
Fig.~\ref{fig:longasym}(c) and the symmetric junction of
Fig.~\ref{fig:longsym}(c) shows that the magnitude of the current
switching due to occupation of a new Andreev level can be nearly the
same for both symmetrical and asymmetrical junctions.  Therefore,
inversion symmetry is not a necessary condition for an Andreev level
to carry a `giant' Josephson current. The ballistic junction ($T=1$)
again carries the largest equilibrium and nonequilibrium Josephson
current.

One further difference between the long SIS junctions with inversion
symmetry and the asymmetric junctions is that the Josephson current
switches by different amounts in the asymmetric junctions when $eV <
\Delta$, as shown in Fig.~\ref{fig:longasym}(b)-(c). In an asymmetric
junction, magnitude of the current $|I_n|$ flowing through each
Andreev level is in general different, while for symmetric junctions
they are nearly the same (as long as $|E_n|$ is not too near the gap
edge). Fig.~\ref{fig:longasym}(b)-(c) and
Fig.~\ref{fig:longsym}(b)-(c) also show that Josephson current does
not fall exactly to zero when probe voltage exceeds the energy gap
($eV \geq \Delta$). As emphasized in Ref.~\cite{chang}, this is
because the leaky Andreev levels outside the superconducting gap carry
a portion of the Josephson current. The normal metal probe therefore
provides a means of doing energy spectroscopy of the Josephson
current, making it possible to measure this `continuum' contribution
to the current. Spectroscopy of the bound levels is again shown
by the differential
conductance $dI_{N}/dV$ in Fig.~\ref{fig:longasym}(d) for
transmission $T = 12\%$.

\twocolumn[\hsize\textwidth\columnwidth\hsize\csname @twocolumnfalse\endcsname

\begin{figure}
\fourfig{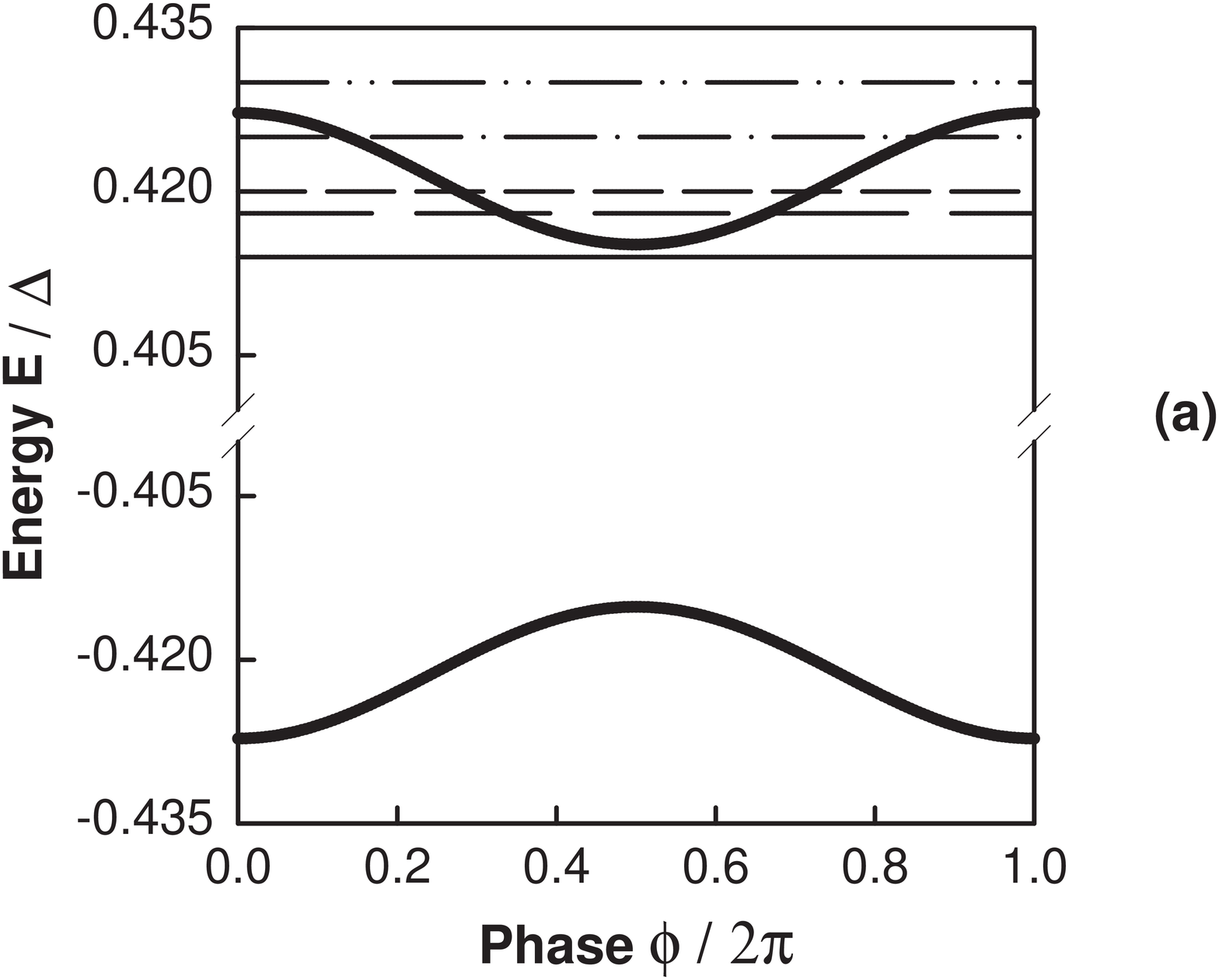}{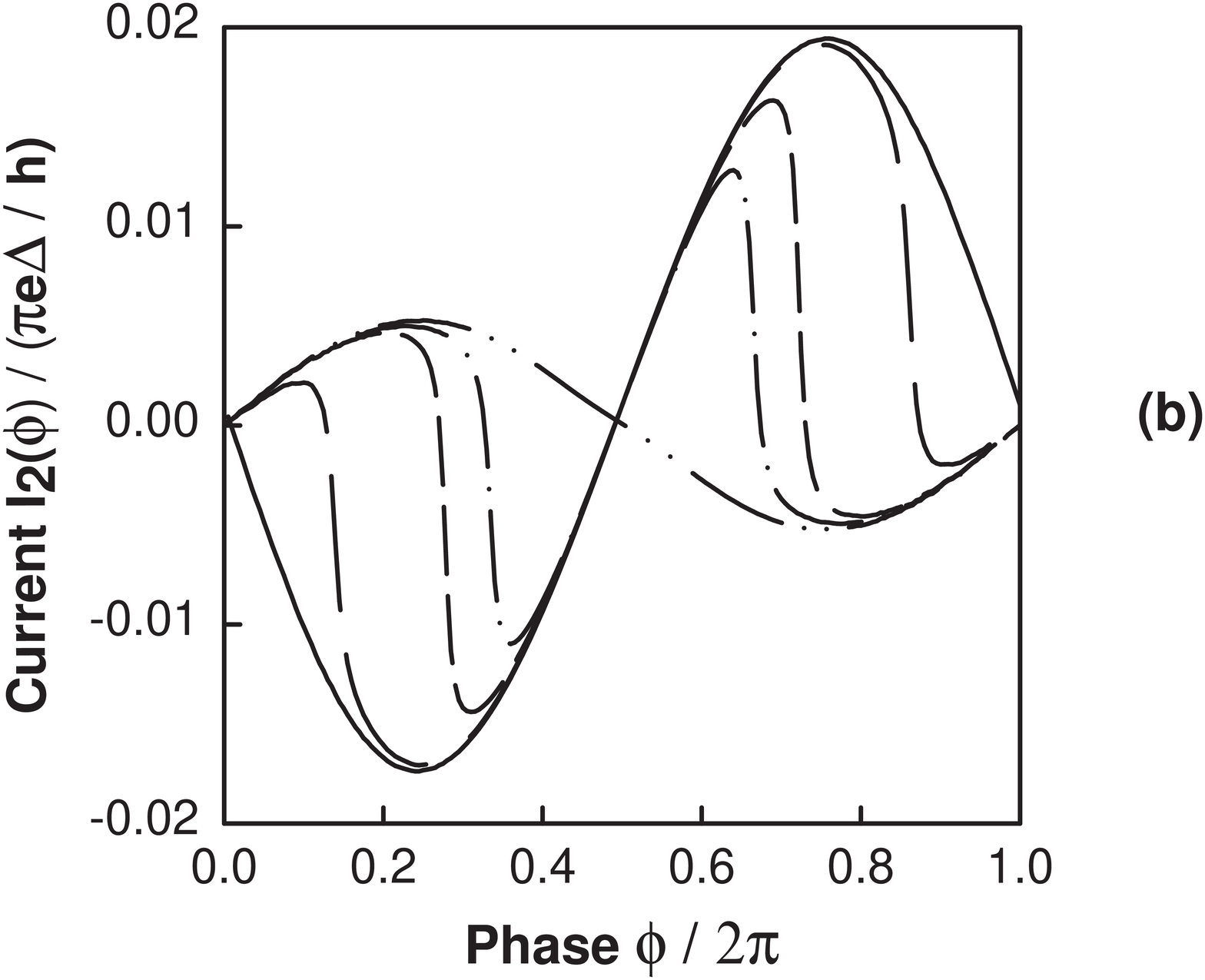}{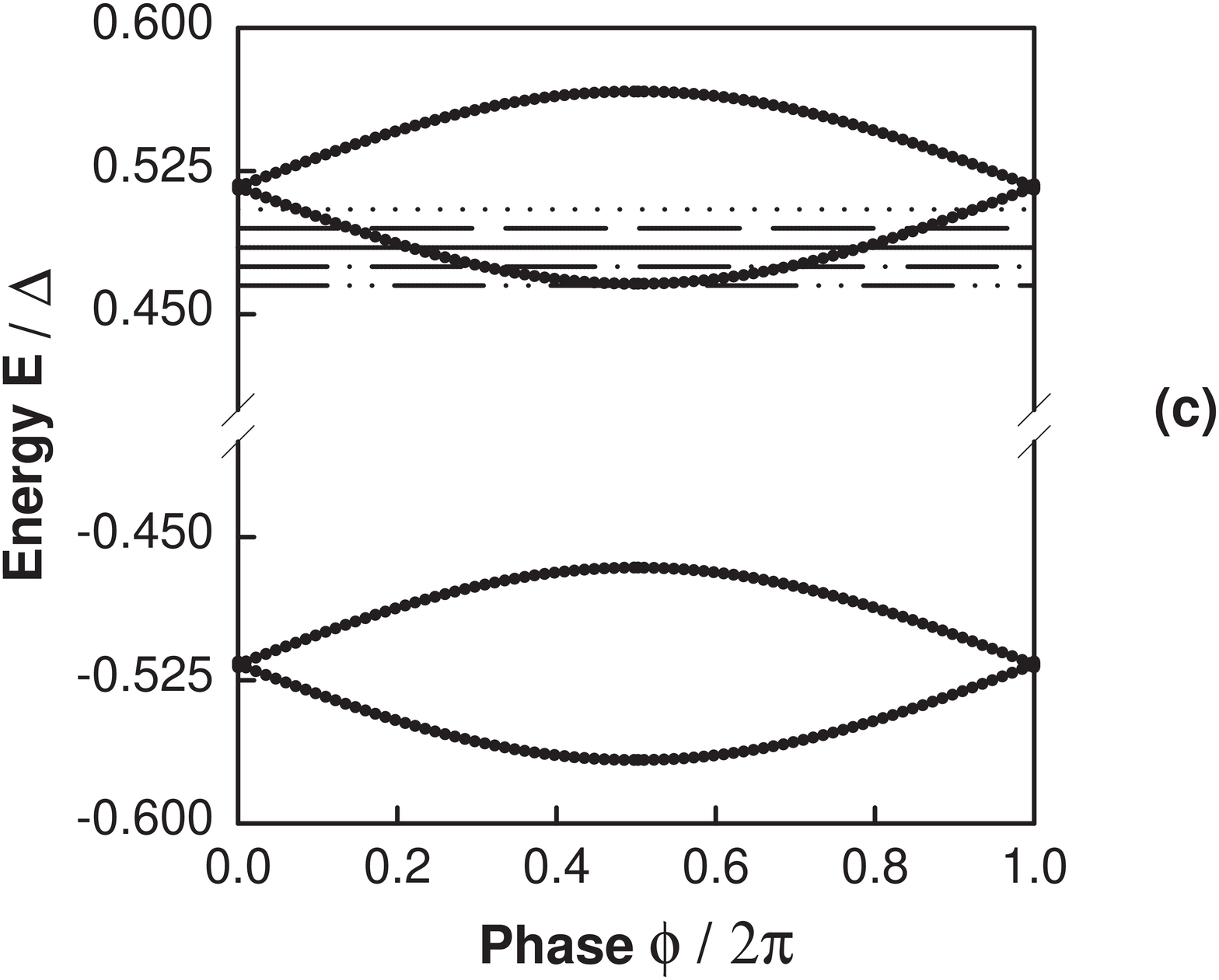}{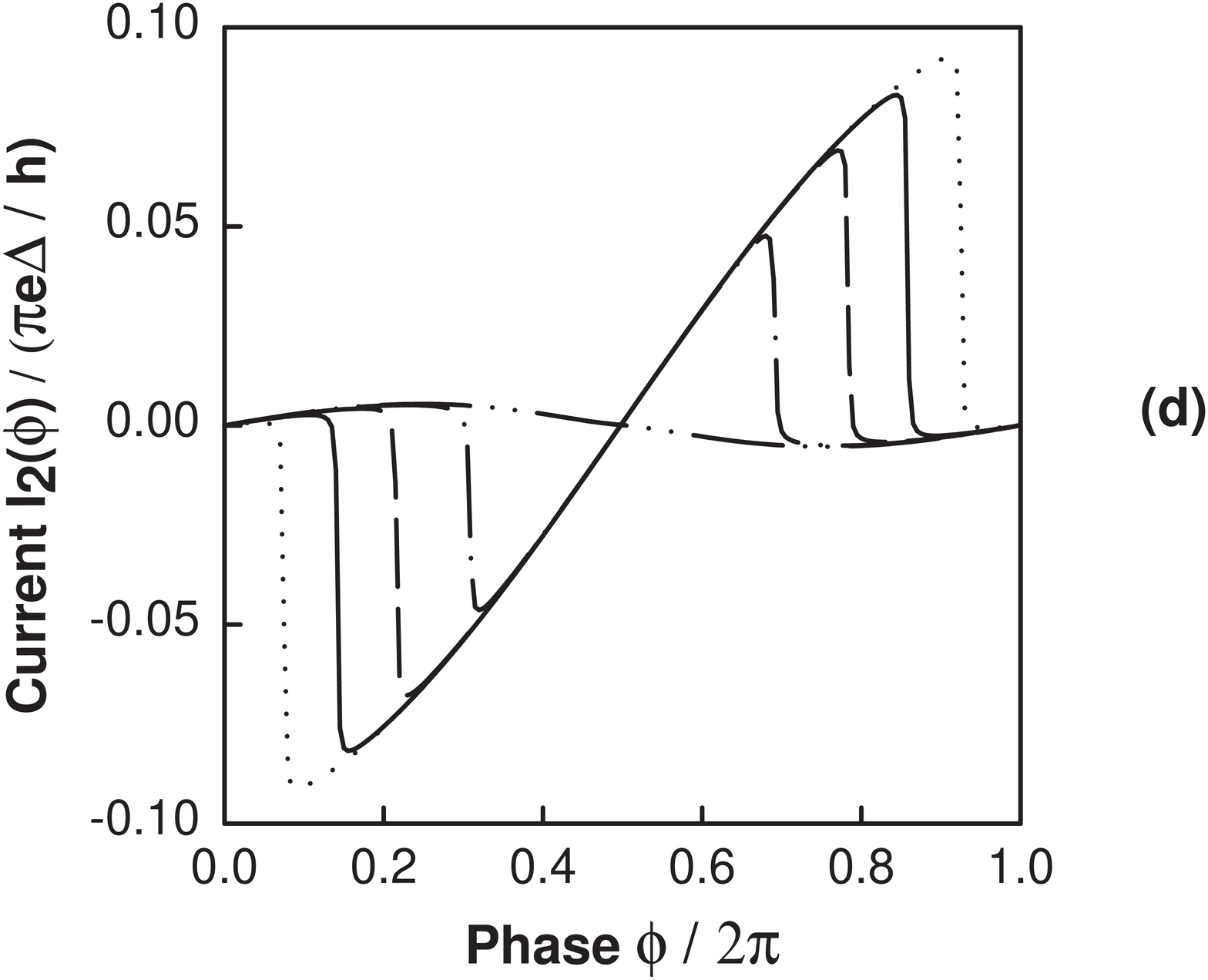}{6.5} 
\caption{Andreev levels $E^{\pm}(\phi)$ 
and nonequilibrium Josephson current-phase relation $I(\phi,V)$
for a junction without inversion symmetry (a)-(b) and with
inversion symmetry (c)-(d). The current phase relations are similar,
except for the larger current magnitude in the symmetric junction.}
\label{fig:longphase}
\end{figure} 

] \narrowtext

\subsection{Current-Phase Relation}
\indent

The current flow through bound Andreev levels can also be observed in
the current-phase relation of a Josephson junction.~\cite{ouboter} In
Fig.~\ref{fig:longphase}, we show the Andreev levels and current-phase
relation in a long $(L = 4 \xi_0)$ SIS junctions.  The asymmetric
junction in (a) and (b) has the impurity placed one third of the
distance across the normal region ($a = L/3$), while the symmetric
junction in (c) and (d) has $a = L/2$.  For both junctions, the
transmission probability $T = 2.5\%$ and the coupling strength
$\epsilon = 0.1\%$.  We apply gate voltages to the SIS junction which
intersects an Andreev level, namely $eV = \pm 0.414, 0.418, 0.420,
0.425, 0.430 \Delta$ in (a) and $eV = \pm 0.465, 0.475, 0.485, 0.495,
0.505 \Delta$ in (c), as shown by the horizontal lines in
Fig.~\ref{fig:longphase}.

Qualitatively, current-phase relation is similar for both types of
Josephson junctions. As the quasi-Fermi energy sweeps through the a
bound Andreev level, the current-phase relation changes from
sinusoidal, to a half-periodic relation, and finally to a $\pi$-phase
shifted junction~\cite{wilhelm}. All these changes in the
current-phase relation are similar to those occurring in ballistic SNS
junctions.~\cite{vwees,chang}, including the half-periodic and
$\pi$-phase shifted (sign change of the Josephson current)
current-phase relations. The current-phase relation for the symmetric
junction evolves from the usual sinusoidal form into a nearly linear
variation of current with phase more typical of ballistic
junctions. The half-periodic current-phase relation can be observed as
a doubling of the AC Josephson frequency~\cite{argaman,zagoskin}.

When we depopulate one of the Andreev levels with
the probe voltage, we increase the magnitude of the critical current.
The equilibrium Josephson current is a small difference between much
larger positive and negative currents flowing in equilibrium.
Depopulating an Andreev level removes some of this current
cancellation in both the asymmetric and symmetric junction types,
increasing the Josephson critical current.  However, the Andreev
levels in the SIS junction with inversion symmetry carry a slightly
larger current (by a factor of $\simeq 4$ in
Fig.~\ref{fig:longphase}). There is a small difference between the
Josephson current betweeen applying positive and negative gate
voltages not shown in Fig.~\ref{fig:longphase}, similar to
Fig.~\ref{fig:shortphase}(b) for the short junction, again due to a
small leakage current from the gate.

\section{Conclusions}
\indent

An additional normal metal terminal weakly coupled to a Josephson
junction, permits one both to determine the Andreev energy levels and
to probe the Josephson current carried through these bound levels.
The differential conductance $dI_N(\phi,V)/dV$ along the normal metal
probe determines the Andreev level positions and width, while changes
in the Josephson current $I(\phi,V)$ as a function of the probe
voltage determine the Andreev level currents. Setting the probe
voltage above the energy gap also allows a measurement of the
`continuum' Josephson current, which flows outside the energy gap. In
this paper we have studied the Josephson current switching and
spectroscopy of the Andreev energy levels in a three terminal SIS
junction, where the normal region of the Josephson junction also
contains an insulator having transmission probability $T \leq 1$. The
results are qualitatively different from the ballistic Josephson
junction ($T = 1$) we considered in an earlier paper~\cite{chang}.

In a short Josephson junction $(L \ll \xi_0)$ containing a tunnel
barrier ($T \ll 1$), the maximum Josephson current switches from the
Ambegaokar-Baratoff value ($e \Delta T / 2 \hbar$) to zero when the
probe voltage is approximately the energy gap, $|eV| \simeq \Delta$.
In a short junction, there are two Andreev levels which carry equal
and opposite currents.  Therefore, populating (or depopulating) both
levels forces the Josephson current to zero. Since the presence of a
tunnel barrier with $T \ll 1$ forces the Andreev levels to the gap
edge at $E \simeq \pm \Delta$, the Josephson current switches to zero
when $|eV| \simeq \Delta$ (independent of the phase
difference). Although the magnitude of the change in the Josephson
current is larger in ballistic junctions (approaching $e\Delta /
\hbar$), the switching voltage in ballistic junctions
ranges between $0 \leq eV \leq \Delta$ (depending on $\phi$).

In a long Josephson junction $(L \gg \xi_0)$ containing a tunnel
barrier ($T \ll 1$), the details of this current switching depend on
the scattering potential inside the normal region. For the long SIS
junction having inversion symmetry, a current proportional to
$\sqrt{T}$ indeed flows through the junction~\cite{wendin} when one
applies certain voltages to the normal metal probe. (The exact current
is $e v_F/ (L + 2\xi(E_n^\pm) ) \sqrt{T}$ where $\xi(E_n^\pm)$ is the
energy dependent coherence length.)  The factor of $\sqrt{T}$ arises
because the inversion symmetry allows degenerate energy levels to
exist even in the presence of a tunnel barrier, similar to the midgap
states in d-wave superconductors.~\cite{riedel-dnd} Although this
nonequilibrium current (proportional to $\sqrt{T}$) is much larger
than the equilibrium Josephson tunneling current (proportional to
$T$), both the changes in the nonequilibrium Josephson current and
magnitude of the equilibrium current are much smaller in SIS junctions
than for ballistic SNS junctions (which approach $e v_F/ (L +
2\xi_0)$.

The bandwidth of the Andreev levels also becomes smaller with
decreasing barrier transmission in long SIS junctions. The range of
gate voltages for which one obtains a nonequilibrium current in the
long SIS junction is equal to this bandwidth, namely $4 \sqrt{T}
\Delta (\xi_0/(L+2\xi_0))$. There are approximately $(L+2\xi_0) / 2
\pi \xi_0$ of these energy levels, so the total range of voltages over
which the Josephson current differs significantly from equilibrium is
$4 \sqrt{T} \Delta / \pi$. Reducing the barrier transmission $T$ to
maximize the size of the nonequilibrium current simultaneously lowers
the range of gate voltages over which one can observe this current. In
a ballistic SNS junction, the Josephson current differs significantly
from its equilibrium value over a much larger range of voltages
$\Delta / 2$, namely half of the energy gap. Nonequilibrium effects on
the Josephson current from the additional normal metal probe are much
larger and occur over a much broader range of gate voltage in
ballistic SNS junctions.

For long SIS junctions which do not possess inversion symmetry, the
energy gaps and currents carried by the Andreev levels are in general
different, so that the Josephson current switches by different amounts
whenever the gate voltage populates a new Andreev level.  The
magnitude of the Josephson current switching can range between the
equilibrium value of the current (proportional to $T$) and the larger
nonequilibrium currents found in symmetric junctions (proportional to
$\sqrt{T}$).  If only a very small energy gap occurs near an Andreev
level crossing, main features of the Josephson current switching and
spectroscopy of the Andreev levels are qualitatively similar to the
symmetric junctions. But because the Andreev levels carry different
currents in general, the cancellation between currents flowing in
opposite directions from two levels adjacent in energy is almost never
exact. The Josephson current can differ from its equilibrium value
over a much wider range of gate voltages in long and asymmetric SIS
junctions.

\section{Acknowledgements}
\indent

We thank Manoj Samanta for many useful discussions. We gratefully
acknowledge support from the MRSEC of the National Science Foundation
under grant No. DMR-9400415 (PFB).

\appendix
\onecolumn
\widetext

\section{Scattering States}
\indent

We compute the Josephson current by finding both the scattering state
and bound state solutions of the BdG equation
\begin{equation}
\left(
        \matrix{
                H(x)-\mu & \Delta(x) \cr
                \Delta^*(x) & -(H^*(x) - \mu)
        }
\right)
\left(
        \matrix{
                u(x) \cr v(x)
        }
\right)
= E
\left(
        \matrix{
                u(x) \cr v(x)
        }
\right)
\;\;\;\;\; .
\label{BdG}
\end{equation}
For our model Josephson junction we take the Hamiltonian
\begin{eqnarray}
H(x) = - \frac{\hbar^2}{2m} \frac{d^2}{dx^2} + V(x) \; ,
\end{eqnarray}
the scattering potential
\begin{eqnarray}
V(x) = V_{\rm s} \delta(x-a) \;\;\;\;\; ,
\label{Vx}
\end{eqnarray}
where $0 \leq a \leq L$, and pair potential
\begin{equation}
\Delta(x) =
\left\{ 
        \matrix{
                \Delta e^{\displaystyle i \phi_1} & x<0 \cr
                0 & 0<x<L \cr
                \Delta e^{\displaystyle i \phi_2} & x>L
        } 
\right. \;\;\;\;\; .
\label{xdelta}
\end{equation} 
We first obtain the solutions of Eq.~(\ref{BdG}) in a uniform
superconductor where $\Delta(x) = {\rm constant}$ and $V(x) = {\rm
constant}$. When the pair potential and electrostatic potential vary
in space, we determine the scattering state solutions of
Eq.~(\ref{BdG}) by matching quasiparticle wave amplitudes at the
potential discontinuities at $x = 0, x = a, x = b$ and $x = L$ as
shown in Fig.~\ref{fig:appfig}. We essentially follow the
calculational methods outlined in the Appendices of
Refs.~\cite{bagwell} and \cite{chang}.

\begin{figure}[htb]
\centps{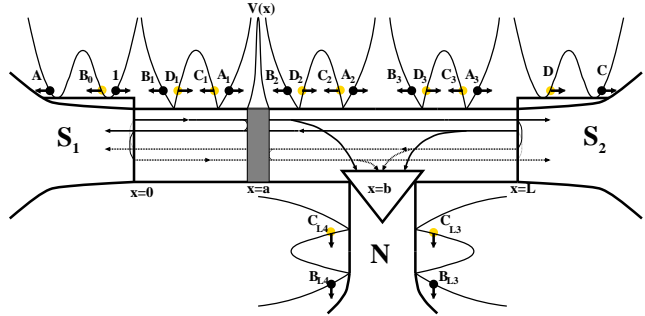}{45}
\setlength{\epsfysize}{7.0truecm}
\caption{An electron-like quasiparticle is injected from the the left 
superconductor reservoir. The wave amplitude described by $B_0$ represents 
the term $B - \frac{u_0}{v_0}$ in Eq.~\ref{ehleft}. The wave amplitudes in the
right superconductor can be obtained interms of $A$ and $B$ by matching the 
boundry conditions at $x = 0, x = a, x = b$ and $x = L$.}
\label{fig:appfig}
\end{figure}

Fig.~\ref{fig:appfig} illusturates an electronlike quasi-particle
injected from the left superconducting reservoir.  In the Andreev
approximation, an electron like excitation incident from the left
superconductor (x \(<\) 0) generates both reflected and transmitted
quasi-particles described by the wave function
\begin{equation}
\left(
      \matrix{
            u_0 e^{\displaystyle i \phi_1} \cr
            v_0
      }
\right)
e^{\displaystyle i k_e x} 
+ \left[
        \matrix{
            B - \frac{u_0}{v_0}
        }
  \right]
\left(
      \matrix{
            v_0 e^{\displaystyle i \phi_1} \cr
            u_0
      }
\right) 
e^{\displaystyle i k_h x}
+ A
\left(
      \matrix{
            u_0 e^{\displaystyle i \phi_1} \cr
            v_0
      }
\right) 
e^{-\displaystyle i k_e x}
\; ,
\label{ehleft}
\end{equation}
valid for ($x \ge 0 \le a$), In Eq.~(\ref{ehleft}), the first term
represents the electronlike quasiparticle injected from the left
superconducting reservoir. The term denoted by the coefficient A
describes the electron-like particle normally reflected from the
impurity while the term denoted by the coefficient B describes the
Andreev reflected hole-like quasi-particle.  The transmitted
electron-like quasiparticle and the Andreev reflected hole-like
quasiparticle in the right superconductor (x \(>\) L) have the wave
function
\begin{equation}
C\left(
      \matrix{
            u_0 e^{\displaystyle i \phi_2} \cr
            v_0
      }
\right) 
e^{\displaystyle i k_e (x-L)} 
+ D
\left(
      \matrix{
            v_0 e^{\displaystyle i \phi_2} \cr
            u_0
      }
\right) 
e^{-\displaystyle i k_h (x-L)}
\; .
\label{eright}
\end{equation}
The term with the coefficient D describes the hole-like normal
reflections from the impurity which are first Andreev reflected from
the right superconductor.  The BCS coherence factors $u_0$ and $v_0$
in Eqs.~(\ref{ehleft})-(\ref{eright}) are
\begin{equation}
2u_0^2= 1 + \frac{\sqrt{E^2-\Delta^2}}{|E|}
\; ,
\label{u0}
\end{equation}
and
\begin{equation}
2v_0^2= 1 - \frac{\sqrt{E^2-\Delta^2}}{|E|}
\; .
\label{v0}
\end{equation}
The wave 
vectors $k_e$ and $k_h$ in Eqs.~(\ref{ehleft})-(\ref{eright}) are
\begin{equation}
k_e = \sqrt{\frac{2m}{\hbar^2}\left( \mu + \sqrt{E^2 - \Delta^2}\right) }
\; ,
\label{ke}
\end{equation}
and
\begin{equation}
k_h = \sqrt{\frac{2m}{\hbar^2}\left( \mu - \sqrt{E^2 - \Delta^2}\right) }
\; .
\label{kh}
\end{equation}
\indent

As shown in Fig.~\ref{fig:appfig}., quasi-particle waves inside the
normal region are grouped into three sets. We obtain the quasiparticle
wave function in the normal regions (b \(<\) x \(<\) L) and (0 \(<\) x
\(<\) a) by matching the waves at pair potential interfaces x = L and
x = 0 respectively.  Therefore, for electrons in region (0 \(<\) x
\(<\) a) we have
\begin{equation}
A_1
\left(
      \matrix{
            1 \cr
            0
      }
\right) 
e^{\displaystyle i \tilde{k}_e x} 
+ C_1 
\left(
      \matrix{
            0 \cr
            1
      }
\right) 
e^{\displaystyle i \tilde{k}_h x} 
=
\left(
      \matrix{
            B v_0 e^{\displaystyle i \phi_1} \cr
            0
      }
\right) 
e^{\displaystyle i \tilde{k}_e x} 
+ 
\left(
      \matrix{
            0 \cr
            (B - \frac{u_0}{v_0} + \frac{v_0}{u_0})u_0
      }
\right) 
e^{\displaystyle i \tilde{k}_h x} 
\; . 
\label{enoa}
\end{equation}
Similarly, in region (b \(<\) x \(<\) L), matching the wave functions yields 
\begin{eqnarray}
A_3
\left(
      \matrix{
            1 \cr
            0
      }
\right) 
e^{\displaystyle i \tilde{k}_e (x-b)} 
+ C_3 
\left(
      \matrix{
            0 \cr
            1
      }
\right) 
e^{\displaystyle i \tilde{k}_h (x-b)} 
 =  \hspace{3.5in} 
\nonumber\\
\left(
      \matrix{
            C u_0 e^{\displaystyle i \phi_2}
                  e^{-\displaystyle i \tilde{k}_e (L-b)} \cr
            0
      }
\right) 
e^{\displaystyle i \tilde{k}_e (x-b)} 
 + \left(
      \matrix{
            0 \cr
            C u_0e^{-\displaystyle i \tilde{k}_h (L-b)}
      }
\right) 
e^{\displaystyle i \tilde{k}_h (x-b)} 
\; . 
\label{enbL}
\end{eqnarray}
For holes in region (0 \(<\) x \(<\) a) we obtain 
\begin{equation}
B_1
\left(
      \matrix{
            1 \cr
            0
      }
\right) 
e^{-\displaystyle i \tilde{k}_e x} 
+ D_1 
\left(
      \matrix{
            0 \cr
            1
      }
\right) 
e^{-\displaystyle i \tilde{k}_h x} 
=
\left(
      \matrix{
            A u_0 e^{\displaystyle i \phi_1} \cr
            0
      }
\right) 
e^{-\displaystyle i \tilde{k}_e x} 
+ 
\left(
      \matrix{
            0 \cr
            Av_0
      }
\right) 
e^{-\displaystyle i \tilde{k}_h x} 
\; ,
\label{hnoa}
\end{equation}
and in region (b \(<\) x \(<\) L) we have
\begin{eqnarray}
B_3
\left(
      \matrix{
            1 \cr
            0
      }
\right) 
e^{-\displaystyle i \tilde{k}_e (x-b)} 
+ D_3 
\left(
      \matrix{
            0 \cr
            1
      }
\right) 
e^{-\displaystyle i \tilde{k}_h (x-b)} 
 = \hspace{3.5in}
\nonumber\\
\left(
      \matrix{
            D v_0 e^{\displaystyle i \phi_2}
                  e^{\displaystyle i \tilde{k}_e (L-b)} \cr
            0
      }
\right) 
e^{-\displaystyle i \tilde{k}_e (x-b)} 
 +
\left(
      \matrix{
            0 \cr
            D u_0e^{\displaystyle i \tilde{k}_h (L-b)}
      }
\right) 
e^{-\displaystyle i \tilde{k}_h (x-b)} 
\; . 
\label{hnbL}
\end{eqnarray}
Since pair potential \(\Delta\) is zero in the normal region, the wave vectors
\(\tilde{k}_e\) and \(\tilde{k}_h\) in Eq.~(\ref{enoa})-(\ref{hnbL}) are 
obtained from Eq.~(\ref{ke}) and (\ref{kh}) by setting \(\Delta\) = 0.

The current amplitudes of incoming and outgoing waves at the impurity
(x = a) and at the normal-metal side probe (x = b) are connected via
two separate scattering matrices.  The scattering matrix for electrons
at x = a is
\begin{equation}
\left(
        \matrix{
        A_2 \cr
        B_1 e^{\displaystyle -i \tilde{k}_e a } 
        } 
\right)
=
\left(
        \matrix{
                t(\tilde{k}_e) & r(\tilde{k}_e) \cr
                r(\tilde{k}_e) & t(\tilde{k}_e) \cr
        } 
\right)
\left(
        \matrix{
        A_1 e^{\displaystyle i \tilde{k}_e a } \cr
        B_2
        } 
\right) \;\;\;\;\; .
\label{esmat} 
\end{equation}
Similarly, for holes at x = a, the scattering matrix is
\begin{equation}
\left(
        \matrix{
        C_1 e^{\displaystyle  \tilde{k}_h a } \cr
        D_2  
        } 
\right)
=
\left(
        \matrix{
                t^*(\tilde{k}_h) & r^*(\tilde{k}_h) \cr
                r^*(\tilde{k}_h) & t^*(\tilde{k}_h) \cr
        } 
\right)
\left(
        \matrix{
        C_2 \cr
        D_1 e^{\displaystyle -i \tilde{k}_h a }
        } 
\right) \;\;\;\;\; .
\label{hsmat} 
\end{equation}
The current transmission and reflection amplitudes for the point
scatterer matrices in Eqs.~(\ref{esmat})-(\ref{hsmat}) are
\begin{equation}
t(k) = \frac{\displaystyle 1}
{\displaystyle 1 + i (m V_{\rm s} /\hbar^2 k)} \;\;\;\;\;
\end{equation}
and
\begin{equation}
r(k) = \frac{\displaystyle - i (m V_{\rm s} /\hbar^2 k)}
{\displaystyle 1 + i (m V_{\rm s} /\hbar^2 k)}
\;\;\;\;\; .
\end{equation}

At $x = b$ the quasiparticle wave amplitudes are coupled by another
scattering matrix to the normal-metal side probe. Since we assumed
that the normal-metal probe only breaks the phase of a quasi-particle,
but not its momentum, we conceptually view the normal probe as two
seperate leads. We couple right-moving quasi-particles only to lead 3
while the left-moving quasi-particles to lead 4. Therefore, the
scattering matrix becomes
\begin{equation}
{\bf S}
=
\left(
        \matrix{
                0 & \sqrt{1-\epsilon} & \sqrt{\epsilon} & 0 \cr
                \sqrt{1-\epsilon} & 0 & 0 & \sqrt{\epsilon} \cr
                \sqrt{\epsilon} & 0 & 0 & -\sqrt{1-\epsilon} \cr
                0 & \sqrt{\epsilon} & -\sqrt{1-\epsilon} & 0 
        } 
\right)
\; .
\label{coupler}
\end{equation}
The scattering matrix ${\bf S}$ from Eq.~(\ref{coupler}) relates the
current amplitudes for electrons in normal region (a \(<\) x \(<\) b)
and normal region (b \(<\) x \(<\) L) as
\begin{equation}
\left(
        \matrix{
                A_3 \cr
       B_2 e^{-\displaystyle i \tilde{k}_e (b-a)} \cr
                B_{L4} \cr
                B_{L3}
        }
\right)
=
\;
         \left( {\bf S} \right)
\;
\left(
        \matrix{
                B_3 \cr
                A_2
        e^{\displaystyle i \tilde{k}_e (b-a)} \cr
                0 \cr
                0
        }
\right)
\; .
\label{econnect}
\end{equation}
The current amplitudes for holes, in the same regions are related as
\begin{equation}
\left(
        \matrix{
                C_2 e^{\displaystyle i \tilde{k}_h (b-a)} \cr
                D_3 \cr
                C_{L4} \cr
                C_{L3} 
        }
\right)
=
\;
         \left( {\bf S} \right)
\;
\left(
        \matrix{
                D_3 e^{-\displaystyle i \tilde{k}_h (b-a)} \cr
                C_3 \cr
                0   \cr
                0
        }
\right)
\; .
\label{hconnect}
\end{equation}

By combining Eqs.~(\ref{ehleft}), (\ref{eright}), (\ref{enoa}),
(\ref{enbL}), (\ref{hnoa}), (\ref{hnbL}), (\ref{esmat}),
(\ref{hsmat}), (\ref{econnect}), and (\ref{hconnect}) we obtain matrix
equations for the coefficients C and D as
\begin{eqnarray}
\left( \frac{v_0}{u_0} \right)
\left(
        \matrix{
                t(\tilde{k}_e) & r(\tilde{k}_e) \cr
                r(\tilde{k}_e) & t(\tilde{k}_e) \cr
        } 
\right)
\left(
        \matrix{
                e^{\displaystyle i (\tilde{k}_e - \tilde{k}_h)a} & 0 \cr
                0 & (1 - \epsilon)  e^{\displaystyle i \phi}
                e^{\displaystyle -i (\tilde{k}_e - \tilde{k}_h)(a-L)} \cr
        } 
\right) \;X \hspace{1.7in}  \nonumber \\ 
\left\{
\left( \frac{v_0}{u_0} \right)
\left(
        \matrix{
                t^*(\tilde{k}_h) & r^*(\tilde{k}_h) \cr
                r^*(\tilde{k}_h) & t^*(\tilde{k}_h) \cr
        } 
\right)
\left(
        \matrix{
                (1 - \epsilon) 
                e^{\displaystyle -i (\tilde{k}_e - \tilde{k}_h)(a-L)} 
                e^{\displaystyle -i \phi}  & 0 \\ 
                0 & e^{\displaystyle i (\tilde{k}_e - \tilde{k}_h)a }  \cr
        } 
\right)
\left(
        \matrix{
                C' \cr A'
        } 
\right) \nonumber \right.\\ 
+
\left.
\left(
        \matrix{
        \left(\frac{\displaystyle u_0}{\displaystyle v_0} -  
                \frac{\displaystyle v_0}{\displaystyle u_0} 
        \right)
        e^{ \displaystyle i \tilde{k}_h a} \cr 0
        } 
\right) 
\right\} 
= \left(
        \matrix{
                1 & 0 \cr
                0 & 1 \cr
        } 
\right)
\left(
        \matrix{
                C' \cr A'
        } 
\right)
\;.
\label{messC}
\end{eqnarray}
and 
\begin{eqnarray}
\left( \frac{v_0}{u_0} \right)^2
\left(
        \matrix{
                t^*(\tilde{k}_h) & r^*(\tilde{k}_h) \cr
                r^*(\tilde{k}_h) & t^*(\tilde{k}_h) \cr
        } 
\right)
\left(
        \matrix{
                (1 - \epsilon)
                e^{\displaystyle -i (\tilde{k}_e - \tilde{k}_h)(a-L)} 
                e^{\displaystyle -i \phi}  & 0 \cr
                0 & e^{\displaystyle i (\tilde{k}_e - \tilde{k}_h)a }  \cr
        } 
\right) \; X \hspace{1.5in} \nonumber \\ 
\left(
        \matrix{
                t(\tilde{k}_e) & r(\tilde{k}_e) \cr
                r(\tilde{k}_e) & t(\tilde{k}_e) \cr
        } 
\right)
\left(
        \matrix{
                e^{\displaystyle i (\tilde{k}_e - \tilde{k}_h)a} & 0 \cr
                0 & (1 - \epsilon)
                e^{\displaystyle -i (\tilde{k}_e - \tilde{k}_h)(a-L)} 
                e^{\displaystyle i \phi} \cr
        } 
\right) 
\left(
        \matrix{
                B' \cr D'
        } 
\right)\;\;\;\;\;\;\;  \nonumber \\ 
+
\left(
        \matrix{
        \left(\frac{\displaystyle u_0}{\displaystyle v_0} -  
                \frac{\displaystyle v_0}{\displaystyle u_0} 
        \right)
        e^{ \displaystyle i \tilde{k}_h a} \cr 0
        } 
\right) \;  
=
\left(
        \matrix{
                1 & 0 \cr
                0 & 1 \cr
        } 
\right)
\left(
        \matrix{ 
                B' \cr D'
        } 
\right)
\;.
\label{messD}
\end{eqnarray}
In Eqs.~(\ref{messC}), the $C'$, $A'$, $B'$, $D'$ are
related to $C$, $A$, $B$, $D$ through
\begin{eqnarray}
C' = \frac{C e^{\displaystyle i \phi}e^{-\displaystyle i \tilde{k}_e (L-a)}}
            {\sqrt{1-\epsilon}} \;\;\;\;\; ,
\label{Cprime}
\end{eqnarray}
\begin{eqnarray}
A' = A e^{-\displaystyle i \tilde{k}_e a} \;\;\;\;\; ,
\label{Aprime}
\end{eqnarray}
\begin{eqnarray}
B' = B e^{\displaystyle i \tilde{k}_h a} \;\;\;\;\; ,
\label{Bprime}
\end{eqnarray}
and
\begin{eqnarray}
D' = \frac{D e^{\displaystyle i \tilde{k}_h (L-a)}}
            {\sqrt{1-\epsilon}} \;\;\;\;\;\; .
\label{Dprime}
\end{eqnarray}

Finally, by inverting the matrix equations (\ref{messC})-(\ref{Dprime}),
we can directly obtain the scattering amplitudes as
\begin{eqnarray}
C_{1 \rightarrow 2}^{e} = t 
\frac{\displaystyle 
        \left( 1 - \left(\frac{v_0}{u_0} \right)^2(1- \epsilon)
        e^{ \displaystyle i \phi } 
        e^{ \displaystyle i (\tilde{k}_e - \tilde{k}_h)L } \right)
        \sqrt{1-\epsilon}
        \left( 1- \left( \frac{v_0}{u_0} \right)^2 \right)
        e^{-\displaystyle i \phi } 
        e^{ \displaystyle i \tilde{k}_e L} 
}
{\displaystyle
        F(\phi, \epsilon, T)
} \; ,
\label{transC}
\end{eqnarray}
and
\begin{eqnarray}
D_{1 \rightarrow 2}^{e} = 
\left(\frac{v_0}{u_0} \right)
\left( 1- \left( \frac{v_0}{u_0} \right)^2 \right) \hspace{3.2in}
\nonumber\\
\frac{\displaystyle 
        \left( t r^*(1 - \epsilon) e^{ \displaystyle -i \phi } 
        e^{ \displaystyle i (\tilde{k}_e - \tilde{k}_h)(L-a) } 
        + r t^* 
        e^{ \displaystyle i (\tilde{k}_e - \tilde{k}_h)a } 
        \right)
        \sqrt{1-\epsilon}
        e^{-\displaystyle i \tilde{k}_h (L-a)} 
        e^{ \displaystyle i \tilde{k}_e a} 
}
{\displaystyle 
        F(\phi, \epsilon, T)
} \; .
\label{transD}
\end{eqnarray}
The resonant denominator $F(\phi, \epsilon, T)$, whose zeroes give the
position and width of the Andreev resonnaces, is
\begin{eqnarray}
\displaystyle
       F(\phi, \epsilon, T)   = 
       1 - 2T( 1 - \epsilon )
       \left( \frac{\displaystyle v_0}{\displaystyle u_0} \right)^2
       e^{ \displaystyle i (\tilde{k}_e - \tilde{k}_h)L } 
       \cos\phi 
       - R
       \left( \frac{\displaystyle v_0}{\displaystyle u_0} \right)^2
       e^{2\displaystyle i (\tilde{k}_e - \tilde{k}_h)a } \;\;\;\; & 
       \nonumber  \\
       - R( 1 - \epsilon )^2
       \left( \frac{\displaystyle v_0}{\displaystyle u_0} \right)^2
       e^{2\displaystyle i (\tilde{k}_e - \tilde{k}_h)(L-a) }
       + ( 1 - \epsilon )^2
       \left( \frac{\displaystyle v_0}{\displaystyle u_0} \right)^4
       e^{2\displaystyle i (\tilde{k}_e - \tilde{k}_h)L } . \;\;\;\;
\end{eqnarray}
For any other quasiparticle, injected from either superconducting
contacts or "conceptual" normal probes, one can obtain the
wave-function amplitudes by calculations similar to the ones described
above. In Appendix B we outline the derivation of the electrical
current flowing into the right superconductor from those wave-function
amplitudes.

\section{Electrical Current}
\indent

The electrical current flow in the right superconductor due to the
injection of a quasiparticle from any lead is~\cite{flow}
\begin{equation}
I_{2} =  \sum_{k;q\beta} 
\left( J_{u} + J_{v} \right)_{q \rightarrow 2}^{ \beta} f_{q}^{ \beta}
-  \sum_{k;q\beta} \left( J_{v} \right)_{q \rightarrow 2}^{ \beta}
\; .
\label{currop}
\end{equation}
We use the indices $q = 1,2,3,4$ as the lead numbers, $k$ as the wave
number of the injected quasi-particle, and $\beta = (e \; {\rm or} \;
h)$ denotes the injection of electron-like or hole-like
quasi-particles. The $J_{u}$ and $J_{v}$ are the Schr\"{o}dinger
currents carried by the waves $u$ and $v$, namely,
\begin{equation}
J_{u}=(e\hbar/m) {\rm Im}\{u^{*}(x) \nabla u(x)\} \; ,
\label{sch_u}
\end{equation}
and
\begin{equation}
J_{v}=(e\hbar/m) {\rm Im}\{v^{*}(x) \nabla v(x)\}.
\label{sch_v}
\end{equation}
The last term in Eq.~(\ref{currop}), $\sum_{\;k;q\beta} \left( J_{v}
\right)_{q \rightarrow 2}^{ \beta}$ = 0, is called `vacuum current' and
is zero in DC problems.  

We can now define the transmission coefficients $T_{q \rightarrow
2}^{\beta}$, which give the ratio of electrical current out to
particle current in, as
\begin{equation}
e v_F T_{q \rightarrow 2}^{\beta} =  
\left(
J_{u} + J_{v} \right)_{q \rightarrow 2}^{ \beta} \; ,
\end{equation}
where $v_F = \hbar k_F / m$. Equation~(\ref{currop}) then reduces to 
\begin{equation}
I_{p} = e \sum_{k;q\beta} v_F T_{q\rightarrow 2}^{ \beta} f_{q}^{ \beta} \;,
\label{landauersup}
\end{equation}
which is the standard Landauer-B\"uttiker form.
The Fermi factor $f_{q}^{ \beta}$ in Eq.~(\ref{landauersup}) is 
\begin{equation}
f_{q}^{ \beta} =  f(E - eV_{q}^{\beta} ) \; ,
\label{fermi}
\end{equation}
where $eV_{q}^{\beta}$ is the effective biasing voltage applied to the
$q$th lead.  Since we ground both superconducting leads, we have
$eV_{1}^{\beta} = eV_{2}^{\beta} =0$. The effective biasing voltages
applied to the normal-metal leads are
\begin{eqnarray}
eV_{3}^{e} = eV_{4}^{e} =  eV , \;\;\;\; \\
eV_{3}^{h} = eV_{4}^{h} =  - eV \; .
\label{vfactors}
\end{eqnarray}

We obtain the current flow into the right superconductor by converting
the sum over $k$ in Eq.~(\ref{landauersup}) into an integral over the
injected energies, namely
\begin{eqnarray}
   I_2(\phi, \epsilon, V, T) & = &
   e ( \int_{-\infty}^{-\Delta} + \int_{\Delta}^{\infty} )
   v_F N_s^+(E) [ T_{1 \rightarrow 2}^{e} + 
                  T_{1 \rightarrow 2}^{h} + 
                  T_{2 \rightarrow 2}^{e} +  
                  T_{2 \rightarrow 2}^{h} ]
        f(E) dE \nonumber \\
   & + & e \int_{-\infty}^{\infty} v_F N_n^+(E) 
   [T_{3 \rightarrow 2}^{e} + 
    T_{4 \rightarrow 2}^{e}] f(E - eV) dE
   \nonumber \\ 
   & + & e \int_{-\infty}^{\infty} v_F N_n^+(E) 
   [T_{3 \rightarrow 2}^{h} + 
    T_{4 \rightarrow 2}^{h}] f(E + eV) dE 
   \; ,
\label{I1itoT}   
\end{eqnarray}
where the superconducting density of states $N_s^+(E)$ related to the
normal density of states $N_n^+(E)$ by
\begin{eqnarray}
N_s^+(E) = \frac{1}{|u_0^2 -  v_0^2|} N_n^+(E)
\; .
\label{NstoNn}
\end{eqnarray}
Combining  the identity $v_F N_s^+(E) \simeq 1/h$, along with
Eqs.~(\ref{I1itoT})-(\ref{NstoNn}), we obtain
\begin{eqnarray}
   I_2(\phi, \epsilon, V, T) & = &
   \frac{e}{h} ( \int_{-\infty}^{-\Delta} + \int_{\Delta}^{\infty} )
                 [T_{1 \rightarrow 2}^{e} + 
                  T_{1 \rightarrow 2}^{h} + 
                  T_{2 \rightarrow 2}^{e} +  
                  T_{2 \rightarrow 2}^{h} ]
        f(E) dE \nonumber \\
   & + & \frac{e}{h} \int_{-\infty}^{\infty} 
   [T_{3 \rightarrow 2}^{e} + 
    T_{4 \rightarrow 2}^{e}] f(E - eV) dE
   \nonumber \\ 
   & + & \frac{e}{h} \int_{-\infty}^{\infty} 
   [T_{3 \rightarrow 2}^{h} + 
    T_{4 \rightarrow 2}^{h}] f(E + eV) dE 
   \; ,
\label{finalI2}   
\end{eqnarray}

The transmission coefficients $T$ in Eq.~(\ref{finalI2}) are related
simply to the scattering amplitudes calculated in Appendix A.  Using
Eq. ~(\ref{sch_u}), (\ref{sch_v}), (\ref{landauersup}) and
(\ref{eright}) we obtain $T_{1 \rightarrow 2}^{e}$, for example, as
\begin{eqnarray}
T_{1 \rightarrow 2}^{e} = 
|C_{1 \rightarrow 2}^{e}|^2 - |D_{1 \rightarrow 2}^{e}|^2 \; .
\label{T1e2}
\end{eqnarray}
The other transmission coefficients and wave-function amplitudes can
be obtained by calculations similar to the one in Appendix A.  Hence
for a hole-like quasiparticle injected from the left superconductor
we have
\begin{eqnarray}
T_{1 \rightarrow 2}^{h} =  |C_{1 \rightarrow 2}^{h}|^2 - 
                           |D_{1 \rightarrow 2}^{h}|^2 \; , 
\label{T1h2}
\end{eqnarray}
where
\begin{eqnarray}
C_{1 \rightarrow 2}^{h} = \left(\frac{v_0}{u_0} \right) \nonumber 
    \left( 1- \left( \frac{v_0}{u_0} \right)^2 \right)  X
    \;\;\;\;\;\;\;\;\;\;\;\;\;\;\;\;\;\;\;\;\;\;\;\;\;\; 
    \;\;\;\;\;\;\;\;\;\;\;\;\;\;\;\;\;\;\;\;\;\;\;\;\;\; 
    \;\;\;\;\;\;\;\;\;\;\;\;\;\;\;\;\;\;\;\;\;\;\;\;\;\;  & \\
\frac{\displaystyle 
        \left( t^* r(1 - \epsilon) e^{ \displaystyle i \phi } 
        e^{ \displaystyle i (\tilde{k}_e - \tilde{k}_h)(L-a) } 
        + r^* t 
        e^{ \displaystyle i (\tilde{k}_e - \tilde{k}_h)a } 
        \right)
        \sqrt{1-\epsilon}
        e^{-\displaystyle i \tilde{k}_h a} 
        e^{-\displaystyle i \phi} 
        e^{\displaystyle i \tilde{k}_e (L-a)} 
}
{\displaystyle 
        F(\phi, \epsilon, T)
} \; ,& 
\label{transC_2}
\end{eqnarray}
and 
\begin{eqnarray}
D_{1 \rightarrow 2}^{h} = t^* 
\frac{\displaystyle 
        \left( 1 - \left(\frac{v_0}{u_0} \right)^2 (1 - \epsilon)
        e^{- \displaystyle i \phi } 
        e^{ \displaystyle i (\tilde{k}_e - \tilde{k}_h)L } \right)
        \left( 1- \left( \frac{v_0}{u_0} \right)^2 \right)
        \sqrt{(1 - \epsilon)}
        e^{-\displaystyle i \tilde{k}_h L} 
}
{\displaystyle
        F(\phi, \epsilon, T)
} \;\; . 
\label{transD_2}
\end{eqnarray}

For an electron-like quasiparticle injected from the right superconductor
we find the transmission probability
\begin{eqnarray}
T_{2 \rightarrow 2}^{e} =  -1 +  
|C_{2 \rightarrow 2}^{e}|^2 
- |D_{2 \rightarrow 2}^{e}|^2 
- (4u_0v_0 - 2\frac{u_0}{v_0}) 
  {\rm Re} \{ D_{2 \rightarrow 2}^{e} \} 
- \frac{u_0^2}{v_0^2} + 4u_0^2 \; ,
\label{T2e2}
\end{eqnarray}
where
\begin{eqnarray}
C_{2 \rightarrow 2}^{e} = r   
\frac{\displaystyle 
        \left( 1 - \left(\frac{v_0}{u_0} \right)^2 
        e^{2\displaystyle i (\tilde{k}_e - \tilde{k}_h)a } \right)
        \left( 1- \left( \frac{v_0}{u_0} \right)^2 \right)
        (1 - \epsilon)
        e^{2\displaystyle i \tilde{k}_e (L-a)} 
}
{\displaystyle
        F(\phi, \epsilon, T)
}  ,\;\;\;\;\;\;\;\;\;\;\;\;\;\;\;\;\;\;\;\;\;\;\;\;\;\;\;\;\;\; 
\label{transC_3}
\end{eqnarray}
and
\begin{eqnarray}
D_{2 \rightarrow 2}^{e} =  
\frac{\displaystyle 
        \left( 1 - \left(\frac{v_0}{u_0} \right)^2 
        [T(1-\epsilon)
        e^{- \displaystyle i \phi } 
        e^{ \displaystyle i (\tilde{k}_e - \tilde{k}_h)L } 
        + 
        R
        e^{2 \displaystyle i (\tilde{k}_e - \tilde{k}_h)a }] \right)
        \left( 1- \left( \frac{v_0}{u_0} \right)^2 \right)
        \left( \frac{u_0}{v_0}  \right)
}
{\displaystyle
        F(\phi, \epsilon, T)
}  . 
\label{transD_3}
\end{eqnarray}

For a hole-like quasiparticle incident from right superconductor,
we find
\begin{eqnarray}
T_{2 \rightarrow 2}^{h} =  1 
-  |C_{2 \rightarrow 2}^{h}|^2 
+ |D_{2 \rightarrow 2}^{h}|^2 + (4u_0v_0 - 2\frac{u_0}{v_0}) 
 {\rm Re} \{D_{2 \rightarrow 2}^{h}\} 
+ \frac{u_0^2}{v_0^2} - 4u_0^2 \; ,
\label{T2h2}
\end{eqnarray}
where
\begin{eqnarray}
C_{2 \rightarrow 2}^{h} =  
\frac{\displaystyle 
        \left( 1 - \left(\frac{v_0}{u_0} \right)^2 
        [T(1-\epsilon)
        e^{ \displaystyle i \phi } 
        e^{ \displaystyle i (\tilde{k}_e - \tilde{k}_h)L } 
        + 
        R
        e^{2 \displaystyle i (\tilde{k}_e - \tilde{k}_h)a }] \right)
        \left( 1- \left( \frac{v_0}{u_0} \right)^2 \right)
        \left( \frac{u_0}{v_0}  \right)
}
{\displaystyle
        F(\phi, \epsilon, T)
}  \; ,
\label{transD_4}
\end{eqnarray}
and
\begin{eqnarray}
D_{2 \rightarrow 2}^{h} = r^*   
\frac{\displaystyle 
        \left( 1 - \left(\frac{v_0}{u_0} \right)^2 
        e^{2\displaystyle i (\tilde{k}_e - \tilde{k}_h)a } \right)
        \left( 1- \left( \frac{v_0}{u_0} \right)^2 \right)
        (1 - \epsilon)
        e^{2\displaystyle i \tilde{k}_h (L-a)} 
}
{\displaystyle
        F(\phi, \epsilon, T)
}  \; .
\label{transC_4}
\end{eqnarray}

For an electron-like quasiparticle injected from the normal probe 
in channel 3 we have
\begin{eqnarray}
T_{3 \rightarrow 2}^{e} =  |C_{3 \rightarrow 2}^{e}|^2 
- |D_{3 \rightarrow 2}^{e}|^2 \; ,
\label{T3e2}
\end{eqnarray}
where
\begin{eqnarray}
C_{3 \rightarrow 2}^{e} = r   
\frac{\displaystyle 
        \left( 1 - \left(\frac{v_0}{u_0} \right)^2 
        e^{2\displaystyle i (\tilde{k}_e - \tilde{k}_h)a } \right)
        \sqrt{|u_0|^2 + |v_0|^2}
        \sqrt{\epsilon(1 - \epsilon)}
        e^{-\displaystyle i \phi_2} 
        e^{\displaystyle i \tilde{k}_h (L+b-2a)} 
}
{\displaystyle
   u_0 F(\phi, \epsilon, T)
}  \; ,
\label{transC_5}
\end{eqnarray}
and
\begin{eqnarray}
D_{3 \rightarrow 2}^{e} =    
        \left( \frac{v_0}{u_0} \right)
        \sqrt{|u_0|^2 + |v_0|^2}
        \sqrt{\epsilon(1 - \epsilon)} 
        e^{-\displaystyle i \phi_2} 
        e^{\displaystyle i \tilde{k}_e (b-a)} 
        e^{-\displaystyle i \tilde{k}_h (L-a)}  
       \;\;\;\;\;\;\;\;\;\;\;\;\;\;\;\;\;\;\;\; 
       \;\;\;\;\;\;\;\;\;\;\;\;\;\;\;\;\;\;\;\; 
       \nonumber \\
\frac{\displaystyle 
        \left( R(1-\epsilon)
        e^{\displaystyle i (\tilde{k}_e - \tilde{k}_h)(L-a) } 
        + 
        T
        e^{\displaystyle i \phi} 
        e^{\displaystyle i (\tilde{k}_e - \tilde{k}_h)a } 
        -
        \left( \frac{v_0}{u_0} \right)^2
        (1-\epsilon) 
        e^{\displaystyle i (\tilde{k}_e - \tilde{k}_h)(L+a) } \right)
}
{\displaystyle
        u_0 F(\phi, \epsilon, T)
} \; .
\label{transD_5}
\end{eqnarray}
For a hole-like quasiparticle incident from 
the normal probe 
in channel 3 we have
\begin{eqnarray}
T_{3 \rightarrow 2}^{h} =  |
D_{3 \rightarrow 2}^{h}|^2 - |C_{3 \rightarrow 2}^{h}|^2 \; , 
\label{T3h2}
\end{eqnarray}
where
\begin{eqnarray}
C_{3 \rightarrow 2}^{h} =    
        \left( \frac{v_0}{u_0} \right)
        \sqrt{|u_0|^2 + |v_0|^2}
        \sqrt{\epsilon(1 - \epsilon)} 
        e^{-\displaystyle i \tilde{k}_h (b-a)} 
        e^{\displaystyle i \tilde{k}_e (L-a)}  
       \;\;\;\;\;\;\;\;\;\;\;\;\;\;\;\;\;\;\;\; 
       \;\;\;\;\;\;\;\;\;\;\;\;\;\;\;\;\;\;\;\; 
       \nonumber \\
\frac{\displaystyle 
        \left( R(1-\epsilon)
        e^{\displaystyle i (\tilde{k}_e - \tilde{k}_h)(L-a) } 
        + 
        T
        e^{-\displaystyle i \phi} 
        e^{\displaystyle i (\tilde{k}_e - \tilde{k}_h)a } 
        -
        \left( \frac{v_0}{u_0} \right)^2
        (1-\epsilon) 
        e^{\displaystyle i (\tilde{k}_e - \tilde{k}_h)(L+a) } \right)
}
{\displaystyle
        u_0 F(\phi, \epsilon, T)
} \; .
\label{transC_6}
\end{eqnarray}
and
\begin{eqnarray}
D_{3 \rightarrow 2}^{h} = r^*   
\frac{\displaystyle 
        \left( 1 - \left(\frac{v_0}{u_0} \right)^2 
        e^{2\displaystyle i (\tilde{k}_e - \tilde{k}_h)a } \right)
        \sqrt{|u_0|^2 + |v_0|^2}
        \sqrt{\epsilon(1 - \epsilon)}
        e^{-\displaystyle i \tilde{k}_h (L+b-2a)} 
}
{\displaystyle
   u_0 F(\phi, \epsilon, T)
}  \; .
\label{transD_6}
\end{eqnarray}

For an electron-like quasiparticle injected from the normal probe in
channel 4 we have
\begin{eqnarray}
T_{4 \rightarrow 2}^{e} =  
|C_{4 \rightarrow 2}^{e}|^2 - |D_{4 \rightarrow 2}^{e}|^2 \; ,
\label{T4e2}
\end{eqnarray}
where
\begin{eqnarray}
C_{4 \rightarrow 2}^{e} =    
        \sqrt{|u_0|^2 + |v_0|^2}
        \sqrt{\epsilon}
        \;\;\;\;\;\;\;\;\;\;\;\;\;\;\;\;\;\;\;\;\;\;\;\;\;
        \;\;\;\;\;\;\;\;\;\;\;\;\;\;\;\;\;\;\;\;\;\;\;\;\;
        \;\;\;\;\;\;\;\;\;\;\;\;\;\;\;\;\;\;\;\;\;\;\;\;\;
        \;\;\;\;\;\;\;\;\;\;\;\;\;\;\;\;\;\;\;\;\;\;\;\;\;
        \nonumber \\
\frac{\displaystyle 
        \left( 1 - \left(\frac{v_0}{u_0} \right)^2 
        [Re^{2\displaystyle i (\tilde{k}_e - \tilde{k}_h)a } 
        + T(1-\epsilon)
        e^{\displaystyle i \phi} 
        Re^{\displaystyle i (\tilde{k}_e - \tilde{k}_h)L } 
        ]
        \right)
        e^{-\displaystyle i \phi_2} 
        e^{\displaystyle i \tilde{k}_e (L-b)}  
}
{\displaystyle
   u_0 F(\phi, \epsilon, T)
}  ,\;\;\;\;\;\;\;\;\;\;\;\;\; 
\label{transC_7}
\end{eqnarray}
and
\begin{eqnarray}
D_{4 \rightarrow 2}^{e} = r^* 
        \left( \frac{v_0}{u_0} \right)
        \sqrt{|u_0|^2 + |v_0|^2}
        \sqrt{\epsilon}
        (1 - \epsilon) 
        \;\;\;\;\;\;\;\;\;\;\;\;\;\;\;\;\;\;\;\;\;\;\;\;\;
        \;\;\;\;\;\;\;\;\;\;\;\;\;\;\;\;\;\;\;\;\;\;\;\;\;
        \;\;\;\;\;\;\;\;\;\;\;\;\;\;\;\;\;\;\;\;\;\;\;\;\;
        \nonumber \\
\frac{\displaystyle 
        \left( 1 - \left(\frac{v_0}{u_0} \right)^2 
        e^{2\displaystyle i (\tilde{k}_e - \tilde{k}_h)a } 
        \right)
        e^{-\displaystyle i \phi_2} 
        e^{\displaystyle i \tilde{k}_e (L-b)} 
        e^{-2\displaystyle i \tilde{k}_h (L-a)}  
}
{\displaystyle
   u_0 F(\phi, \epsilon, T)
}   . \;\;\;\;\;\;\;\;\;\;\;\;\; 
\label{transD_7}
\end{eqnarray}

For a hole-like quasiparticle incident 
from the normal probe in
channel 4 we have
\begin{eqnarray}
T_{4 \rightarrow 2}^{h} =  
|C_{4 \rightarrow 2}^{h}|^2 - |D_{4 \rightarrow 2}^{h}|^2 \; ,
\label{T4h2}
\end{eqnarray}
where
\begin{eqnarray}
C_{4 \rightarrow 2}^{h} = r 
        \left( \frac{v_0}{u_0} \right)
        \sqrt{|u_0|^2 + |v_0|^2}
        \sqrt{\epsilon}
        (1 - \epsilon) 
        \;\;\;\;\;\;\;\;\;\;\;\;\;\;\;\;\;\;\;\;\;\;\;\;\;
        \;\;\;\;\;\;\;\;\;\;\;\;\;\;\;\;\;\;\;\;\;\;\;\;\;
        \;\;\;\;\;\;\;\;\;\;\;\;\;\;\;\;\;\;\;\;\;\;\;\;\;
        \nonumber \\
\frac{\displaystyle 
        \left( 1 - \left(\frac{v_0}{u_0} \right)^2 
        e^{2\displaystyle i (\tilde{k}_e - \tilde{k}_h)a } 
        \right)
        e^{-\displaystyle i \tilde{k}_h (L-b)} 
        e^{2\displaystyle i \tilde{k}_e (L-a)}  
}
{\displaystyle
   u_0 F(\phi, \epsilon, T)
}  ,\;\;\;\;\;\;\;\;\;\;\;\;\; 
\label{transC_8}
\end{eqnarray}
and
\begin{eqnarray}
D_{4 \rightarrow 2}^{h} =    
        \sqrt{|u_0|^2 + |v_0|^2}
        \sqrt{\epsilon}
        \;\;\;\;\;\;\;\;\;\;\;\;\;\;\;\;\;\;\;\;\;\;\;\;\;
        \;\;\;\;\;\;\;\;\;\;\;\;\;\;\;\;\;\;\;\;\;\;\;\;\;
        \;\;\;\;\;\;\;\;\;\;\;\;\;\;\;\;\;\;\;\;\;\;\;\;\;
        \;\;\;\;\;\;\;\;\;\;\;\;\;\;\;\;\;\;\;\;\;\;\;\;\;
        \nonumber \\
\frac{\displaystyle 
        \left( 1 - \left(\frac{v_0}{u_0} \right)^2 
        [Re^{2\displaystyle i (\tilde{k}_e - \tilde{k}_h)a } 
        + T(1-\epsilon)
        e^{-\displaystyle i \phi} 
        e^{\displaystyle i (\tilde{k}_e - \tilde{k}_h)L } 
        ]
        \right)
        e^{-\displaystyle i \tilde{k}_h (L-b)}  
}
{\displaystyle
   u_0 F(\phi, \epsilon, T)
}  .
\label{transD_8}
\end{eqnarray}

\end{document}